\documentclass[onecolumn,12pt,twoside]{IEEEtran}
\ifCLASSINFOpdf
 \usepackage[pdftex]{graphicx}
\else
  \usepackage[dvips]{graphicx}
\fi

\usepackage[cmex10]{amsmath}
\usepackage{amsfonts}
\usepackage{amssymb}
\usepackage{amsthm}
\usepackage{morefloats}
\usepackage{cite}
\usepackage{bm}
\usepackage{stfloats}

\usepackage[caption=false,font=normalsize,labelfont=sf,textfont=sf]{subfig}
\begin{document}
%
\title{Asymptotic Rate Analysis of Downlink Multi-user Systems with Co-located and Distributed Antennas \thanks{This paper was presented in part at the IEEE Wireless Communications and Networking Conference (WCNC), Shanghai, Apr. 2013.}}

\author{Junyuan~Wang,~\IEEEmembership{Student Member,~IEEE},~and~Lin~Dai,~\IEEEmembership{Senior Member,~IEEE} \thanks{J. Wang and L. Dai are with the Department of Electronic Engineering, City University of Hong Kong, 83 Tat Chee Avenue, Kowloon Tong, Hong Kong, China (email: jywang6-c@my.cityu.edu.hk; lindai@cityu.edu.hk). }}

\maketitle

\begin{abstract}
A great deal of efforts have been made on the performance evaluation of distributed antenna systems (DASs). Most of them assume a regular base-station (BS) antenna layout where the number of BS antennas is usually small. With the growing interest in cellular systems with large antenna arrays at BSs, it becomes increasingly important for us to study how the BS antenna layout affects the rate performance when a massive number of BS antennas are employed.

This paper presents a comparative study of the asymptotic rate performance of downlink multi-user systems with multiple BS antennas either co-located or uniformly distributed within a circular cell. Two representative linear precoding schemes, maximum ratio transmission (MRT) and zero-forcing beamforming (ZFBF), are considered, with which the effect of BS antenna layout on the rate performance is characterized. The analysis shows that as the number of BS antennas $L$ and the number of users $K$ grow infinitely while $L/K{\rightarrow}\upsilon$, the asymptotic average user rates with the co-located antenna (CA) layout for both MRT and ZFBF are logarithmic functions of the ratio $\upsilon$. With the distributed antenna (DA) layout, in contrast, the scaling behavior of the average user rate closely depends on the precoding schemes. With ZFBF, for instance, the average user rate grows unboundedly as $L, K{\rightarrow} \infty$ and $L/K{\rightarrow}\upsilon{>}1$, which indicates that substantial rate gains over the CA layout can be achieved when the number of BS antennas $L$ is large. The gain, nevertheless, becomes marginal when MRT is adopted.
\end{abstract}



\section{Introduction}
The distributed antenna system (DAS) has become a promising candidate for future mobile communication systems thanks to its open architecture and flexible resource management \cite{LTE,Heath_overview}. In DASs, many remote antenna ports are geographically distributed over a large area and connected to a central processor by fiber or coaxial cable. Although the idea of DAS was originally proposed to cover the dead spots in indoor wireless communication systems \cite{Saleh}, research activities on cellular DASs have been intensified in the past few years owing to the fast growing demand for high data-rate services \cite{Hu,WirelessCom,JSAC}.

For cellular systems, the use of distributed base-station (BS) antennas enables efficient utilization of spatial resources, which, on the other hand, also significantly complicates the channel modeling and system analysis. In contrast to the classical point-to-point multiple-input-multiple-output (MIMO) channel, paths between the distributed BS antennas and each single user are subject to different levels of large-scale fading, which are sensitive to the positions of BS antennas and users. To obtain the average ergodic capacity of a single-user DAS (i.e., the ergodic capacity is further averaged over the large-scale fading coefficients), a symmetric distributed MIMO channel was assumed in \cite{Roh,HDai} where the user has equal access distance to each distributed BS antenna but the shadowing coefficients independently follow the log-normal distribution. The effect of path loss was further incorporated into the channel model in \cite{Zhuang,Xiao,Choi,Wang,Feng1,Zhu,Lee}. Most of them assume a regular BS antenna layout which, as pointed out in \cite{Zhuang}, may be difficult to implement in practice due to complicated geographic conditions especially when the number of BS antennas is large. A random antenna layout, in contrast, describes a more general scenario and provides a reasonable performance lower-bound. With BS antennas uniformly distributed over a circular area, the distribution of access distance was characterized in \cite{Dai_JSAC}, and shown to be crucially determined by the total number of BS antennas.

In the multi-user scenario, the capacity region of downlink multi-user MIMO in Gaussian channels has been characterized in \cite{Weingarten}, which showed that the optimal precoding scheme is a pre-interference-cancelation strategy known as dirty-paper coding (DPC) \cite{Costa}. Despite the information-theoretical optimality, it is difficult to implement DPC in practice. A number of linear precoding schemes were, therefore, proposed to trade off between rate performance and system complexity (see\cite{Gesbert} and the references therein). Among them, maximum ratio transmission (MRT) \cite{MRT} and zero-forcing beamforming (ZFBF) \cite{Caire} are two representative non-orthogonal and orthogonal precoding schemes, respectively.

For downlink multi-user DASs, the linear precoding schemes developed for the traditional multi-user MIMO systems can be applied in a straightforward manner. Compared to the single-user case, however, much fewer studies focused on the performance evaluation of downlink multi-user DASs \cite{Park,Ahmad,Kim,Heath,Huang}. The difficulty originates from the fact that with distributed antennas, the rate performance is crucially determined by the positions of BS antennas and users. When the number of BS antennas is large, the numerical calculation of the downlink user rate may require prohibitively high complexity. For computational tractability, various simplified transmission schemes have been proposed \cite{Park,Kim}, and a regular BS antenna layout with a small number of BS antennas is usually assumed \cite{Park,Ahmad,Kim,Heath,Huang}.

Recently, there has been a growing interest in cellular systems with large antenna arrays at BSs \cite{LargeMIMO_Overview,JSAC_LargeMIMO}. With hundreds of antennas employed at the BS side, even the performance evaluation of the traditional multi-user MIMO systems becomes challenging. In this scenario, the asymptotic analysis proves to be a useful tool: the limiting behavior of large-scale systems normally becomes deterministic and leads to simple explicit expressions which provide good approximations for the finite case and shed important light on the practical system design.

The asymptotic analysis has been widely adopted in traditional MIMO systems \cite{Lozano,Tulino_1,Hoydis,Matthaiou} by applying asymptotic results from random matrix theory \cite{Tulino,Couillet}. For multi-user DASs, the asymptotic uplink sum capacity with $L$ BS antenna clusters and $K$ users was derived in \cite{Aktas,Zhang} as an implicit function of $L\times K$ large-scale fading coefficients by assuming that the number of antennas in each BS antenna cluster and the number of user antennas go to infinity while their ratio is fixed. The computational complexity, however, sharply increases with $L$ and $K$, which makes it difficult to analyze the effect of BS antenna layout on the sum capacity when a large number of users and BS antennas are distributed in the area. To compare the uplink sum capacity of DASs to that of cellular systems with co-located BS antennas for large $L$ and $K$, asymptotic bounds were further developed in \cite{Dai_JSAC}. The analysis showed that substantial capacity gains achieved by the DAS mainly come from 1) the reduction of the minimum access distance of each user; and 2) the enhanced channel fluctuation. In the downlink, how the BS antenna layout affects the rate performance further depends on the precoding schemes, which remains largely unknown.

In this paper, an asymptotic rate analysis is presented for a downlink multi-user system where $K$ single-antenna users are uniformly distributed and $L$ BS antennas are either co-located or uniformly distributed within a circular cell. For demonstration, two representative non-orthogonal and orthogonal linear precoding schemes, MRT \cite{MRT} and ZFBF \cite{Caire}, are considered. As $L, K{\rightarrow} \infty$ and $L/K{\rightarrow} \upsilon$,\footnote{It should be distinguished from previous asymptotic analysis \cite{Aktas, Zhang} where the number of antennas in each BS antenna cluster and the number of user antennas grow infinitely, while the number of BS antenna clusters $L$ and the number of users $K$ are finite, and usually small for computational tractability.} the asymptotic average user rates with the co-located antenna (CA) layout are derived and shown to be good approximations for the finite case. With the distributed antenna (DA) layout, bounds are developed to study the scaling behavior of the rate performance.

Our analysis shows that with MRT, the maximum achievable ergodic rate of each user in the CA layout is solely determined by the ratio of the number of BS antennas $L$ and the number of users $K$. With the DA layout, in contrast, the intra-cell interference varies with the BS antenna topology, and the ergodic rate of each user becomes dependent on the positions of BS antennas and users. To characterize the scaling behavior of the average user rate with the DA layout, an asymptotic upper-bound is further obtained as $L,K\to\infty$ and $L/K\to\upsilon$. Both the asymptotic average user rate in the CA layout and the asymptotic upper-bound of the average user rate in the DA layout are found to be logarithmically increasing with $\upsilon$, but in the orders of $\log_2 \upsilon$ and $\frac{\alpha}{2}\log_2 \upsilon$, respectively, where $\alpha>2$ denotes the path-loss factor. 

For each user in the DA layout, both the desired signal power and the intra-cell interference are significantly enhanced owing to the reduction of the minimum access distance to BS antennas. If an orthogonal precoding scheme is adopted such that the intra-cell interference is eliminated by joint precoding among users, more prominent rate gains can be expected in the DA case. The analysis corroborates that with ZFBF, the average user rate in the DA layout grows with the number of BS antennas $L$ in the order of $\log_{2}\left(\left(L-K+1\right)^{\frac{\alpha}{2}}/K\right)$, which becomes infinite as $L, K{\rightarrow} \infty$ and $L/K{\rightarrow} \upsilon > 1$. It is in sharp contrast to the CA layout where the asymptotic average user rate is a logarithmic function of $\upsilon$. Substantial rate gains over the CA layout can be achieved when the number of BS antennas $L$ is large.


The remainder of the paper is organized as follows. Section II introduces the system model. The asymptotic rate analysis with MRT and ZFBF is presented in Section III and Section IV, respectively. Concluding remarks are summarized in Section V.

Throughout this paper, the superscript $\dag$ denotes conjugate transpose. $\mathbb{E}[\cdot]$ denotes the expectation operator. $\|\mathbf{x}\|$ denotes the Euclidean norm of vector $\mathbf{x}$. $\mathbf{I}_{K}$ denotes a $K\times K$ identity matrix. $\mathbf{1}_{1\times L} $ denotes a $1 \times L$ matrix with all entries one. $\det(\mathbf{X})$ denotes the determinant of matrix $\mathbf{X}$. $x\sim \mathcal{CN}(u,\sigma^2)$ denotes a complex Gaussian random variable with mean $u$ and variance $\sigma^2$. $\mathbf{X}\sim \mathcal{W}_K(L,\mathbf{\Sigma})$ denotes a $K\times K$ Wishart matrix $\mathbf{X}$ with $L$ degrees of freedom and covariance matrix $\mathbf{\Sigma}$. $|\it{\mathcal{X}}|$ denotes the cardinality of set $\it{\mathcal{X}}$.

\vspace{5mm}
\section{System Model}
Consider the downlink transmission of a multi-user system with a set of users, denoted by $\mathcal{K}$, and a set of base-station (BS) antennas, denoted by $\mathcal{B}$, with $|\mathcal{K}|=K$ and $|\mathcal{B}|=L$. Suppose that $K$ users are uniformly distributed within a circular cell, and each user is equipped with a single antenna. Without loss of generality, the radius of the circular cell is normalized to be 1.

Let us focus on the downlink performance of the $k$th user. The received signal of user $k$ can be written as
\begin{equation}\label{signal model}
y_{k}=\underbrace{\mathbf{g}_{k}\mathbf{x}_{k}}_{\text{desired signal}}
+\underbrace{\sum_{j\in \mathcal{K}, j \neq k}\mathbf{g}_{k}\mathbf{x}_{j}}_{\text{intra-cell interference}}+z_{k},
\end{equation}
where $\mathbf{x}_{k}{\in} \mathbb{C}^{L\times 1}$ is the signal transmitted from the BS to user $k$. $z_{k}\sim \mathcal{CN}(0,N_0)$ is the additive white Gaussian noise (AWGN) at user $k$. $\mathbf{g}_{k}{\in} \mathbb{C}^{1\times L}$ is the channel gain vector from the BS to user $k$, which can be written as
\begin{equation}\label{define of g}
\mathbf{g}_{k}=\bm{\gamma}_{k}\circ\mathbf{h}_{k},
\end{equation}
where $\mathbf{h}_{k}{\in}\mathbb{C}^{1\times L}$ denotes the small-scale fading vector with entries modeled as independent and identically distributed (i.i.d) complex Gaussian random variables with zero mean and unit variance. $\bm{\gamma}_{k}{\in}\mathbb{R}^{1\times L}$ is the large-scale fading vector from the BS to user $k$.  $\circ$ represents the Hadamard product.

Moreover, we assume that full channel state information (CSI) is perfectly known at both the transmitter side and the receiver side. With linear precoding, the transmitted signal for user $j$ can be written as
\begin{equation}\label{define of x}
\mathbf{x}_{j}=\mathbf{w}_{j} \cdot s_{j},
\end{equation}
for any $j\in\mathcal{K}$, where $s_{j}\sim \mathcal{CN}(0, \bar{P}_{j})$ is the information-bearing signal and $\mathbf{w}_{j}$ is the precoding vector with $\|\mathbf{w}_{j}\|=1$. The total transmission power of the BS is assumed to be fixed at $P_{t}$, and $P_t$ is equally allocated to each user with
\begin{equation}\label{power per user}
\bar{P}_{j}=\frac{P_t}{K},
\end{equation}
$j\in\mathcal{K}$. The second term on the right-hand side of (\ref{signal model}), i.e., $u_{k}^{intra}=\sum_{j\in\mathcal{K}, j \neq k}\mathbf{g}_{k}\mathbf{x}_{j}$, denotes the intra-cell interference received at user $k$. With a large number of BS antennas, $u_{k}^{intra}$ can be modeled as a complex Gaussian random variable with zero mean and variance $I_{k}^{intra}$. It can be easily obtained from (\ref{define of x}) that
\begin{equation}\label{define of I_intra}
\begin{aligned}
I_{k}^{intra}&=\sum_{j\in\mathcal{K}, j\neq k}\mathbb{E}\left[\mathbf{g}_{k}\mathbf{w}_{j}
\mathbf{w}_{j}^{\dag}\mathbf{g}_{k}^{\dag}\right] \bar{P}_{j}.
\end{aligned}
\end{equation}

In this paper, we normalize the total system bandwidth into unity and focus on the spectral efficiency. The maximum achievable ergodic rate of user $k\in\mathcal{K}$ can be written as
\begin{equation}\label{define of Rk}
R_{k}=\mathbb{E}_{\mathbf{h}_{k}}\left[\log_{2}\left(1+
\frac{\bar{P}_{k}\|\bm{\gamma}_{k}\|^{2}\tilde{\mathbf{g}}_{k}\mathbf{w}_{k}\mathbf{w}_{k}^{\dag}\tilde{\mathbf{g}}_{k}^{\dag}}
{N_{0}+I_{k}^{intra}}\right)\right],
\end{equation}
where
\begin{equation}\label{nomarlized g}
\tilde{\mathbf{g}}_{k}=\bm{\beta}_{k}\circ\mathbf{h}_{k}
\end{equation}
denotes the normalized channel gain vector from the BS to user $k$, and $\bm{\beta}_{k}$ is the normalized large-scale fading vector with entries
\begin{equation}\label{define of beta}
\beta_{k, l}=\frac{\gamma_{k,l}}{\|\bm{\gamma}_{k}\|},
\end{equation}
$l\in\mathcal{B}$. Obviously, we have $\sum_{l\in\mathcal{B}}\beta_{k,l}^{2}=1$ for any user $k\in\mathcal{K}$. Let
\begin{equation}\label{define of mu}
\mu_{k}=\frac{\bar{P}_{k}\|\bm{\gamma}_{k}\|^{2}}{N_{0}+I_{k}^{intra}}
\end{equation}
denote the average received signal-to-interference-plus-noise ratio (SINR) of user $k$. By substituting (\ref{define of mu}) into (\ref{define of Rk}), the maximum achievable ergodic rate $R_{k}$ can be further written as
\begin{equation}\label{Rk_mu}
R_{k}=\mathbb{E}_{\mathbf{h}_{k}}\left[\log_{2}\left(1+\mu_{k}
\tilde{\mathbf{g}}_{k}\mathbf{w}_{k}\mathbf{w}_{k}^{\dag}\tilde{\mathbf{g}}_{k}^{\dag}\right)\right].
\end{equation}

\begin{figure}[t]
\begin{center}
\includegraphics[width=0.75\textwidth]{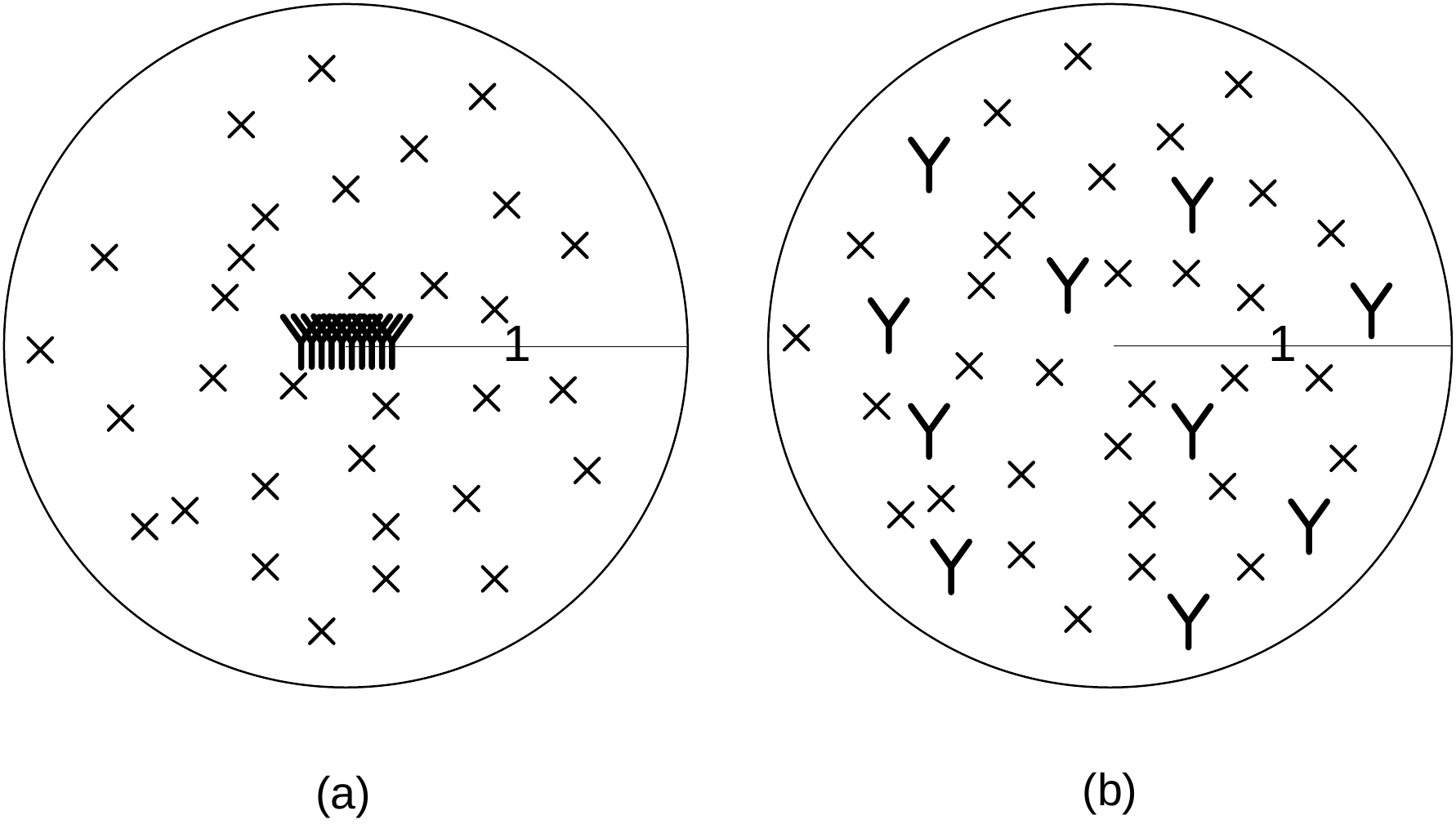}
\caption{Two BS antenna layouts are considered in this paper. (a) With the CA layout, all the BS antennas are co-located at the center of the cell. (b) With the DA layout, the BS antennas are uniformly distributed within the cell. "x" represents a user and "Y" represents a BS antenna.}
\label{FIG_SingleCell}
\end{center}
\end{figure}

It is clear from (\ref{nomarlized g}) and (\ref{Rk_mu}) that the ergodic rate of user $k$ closely depends on the normalized large-scale fading vector $\bm{\beta}_{k}$, which varies with the BS antenna topology and user $k$'s position. In this paper, we consider two BS antenna layouts as shown in Fig. \ref{FIG_SingleCell}: (a) the co-located antenna (CA) layout where all the BS antennas are placed at the center of the circular cell, and (b) the distributed antenna (DA) layout where the BS antennas are uniformly distributed within the cell.\footnote{Note that a regular BS antenna layout is usually assumed in most of previous studies \cite{Choi, Wang, Zhu, Lee, Park, Kim, Ahmad, Heath, Huang}. As we point out in Section I, such a regular BS antenna layout may be difficult to implement in practice when the number of BS antennas is large. Therefore, in this paper, we consider a random BS antenna layout which can also better describe the geographical randomness.} Without loss of generality, we ignore the shadowing effect and model the large-scale fading coefficient as
\begin{equation}\label{define of gamma}
\gamma_{k,l}=\|\mathbf{r}_{l}^{B}-\mathbf{r}_{k}^{U}\|^{-\frac{\alpha}{2}},
\end{equation}
where $\alpha$ is the path-loss factor. $\mathbf{r}_{l}^{B}$ is the position of the $l$th BS antenna and $\mathbf{r}_{k}^{U}$ is the position of user $k$. With the CA layout, all the BS antennas are co-located at the center of the circular cell, i.e., $\mathbf{r}_{l}^{B}=(0, 0)$ for any $l\in\mathcal{B}$. As a result, for any user $k\in \mathcal{K}$, the large-scale fading coefficients to $L$ BS antennas are identical, i.e., $\gamma_{k,1}=...=\gamma_{k,L}$. The normalized large-scale fading vector is then given by $\bm{\beta}_{k}=\frac{1}{\sqrt{L}}\mathbf{1}_{1\times L}$, according to (\ref{define of beta}). Different from the CA case, with the DA layout, the normalized large-scale fading vector $\bm{\beta}_{k}$ has unequal entries as user $k$ has different access distances to $L$ BS antennas.

Because $\bm{\beta}_{k}$ is determined by the positions of user $k$ and $L$ BS antennas, we further define the average maximum achievable ergodic rate of users (which is referred to as "average user rate" in the following) as
\begin{equation}\label{define of AverageR}
\bar{R}\triangleq\mathbb{E}_{\left\{\mathbf{r}_{k}^{U}\right\}_{k\in\mathcal{K}},\left\{\mathbf{r}_{l}^{B}\right\}_{l\in\mathcal{B}}}\left[R_{k}\right],
\end{equation}
where the maximum achievable ergodic rate of user $k$, $R_{k}$, is averaged over all possible positions of BS antennas and users. Note that with the CA layout, $\mathbf{r}_{l}^{B}$ is fixed at $(0, 0)$ for any BS antenna $l\in\mathcal{B}$. The average user rate is thus reduced to
\begin{equation}\label{define of AverageR_CA}
\bar{R}^{C}\triangleq\mathbb{E}_{\left\{\mathbf{r}_{k}^{U}\right\}_{k\in\mathcal{K}}}\left[R_{k}^{C}\right].
\end{equation}

In addition to the BS antenna topology, we can see from (\ref{Rk_mu}) that the rate performance also crucially depends on the precoding vector $\mathbf{w}_{k}$. In the following sections, we will focus on two representative non-orthogonal and orthogonal linear precoding schemes: maximum ratio transmission (MRT) \cite{MRT} and zero-forcing beamforming (ZFBF) \cite{Caire}. We are particularly interested in the comparison of the rate performance with the CA and DA layouts under different precoding schemes when the number of BS antennas $L$ and the number of users $K$ are both large.

\section{Rate Analysis with MRT}
MRT is a representative non-orthogonal linear precoding scheme, with which the precoding vector is given by \cite{MRT}
\begin{equation}\label{define of w_M}
\mathbf{w}_{j}=\frac{{\mathbf{g}_{j }}^{\dag}}{\|\mathbf{g}_{j }\|},
\end{equation}
for any user $j\in\mathcal{K}$. It can be easily obtained by combining (\ref{define of I_intra}) and (\ref{define of w_M}) that
\begin{equation}\label{I_intra}
\begin{aligned}
I_{k}^{intra}=\sum_{j\in\mathcal{K}, j\neq k}\sum_{l\in\mathcal{B}}a_{j,l} \cdot \gamma_{k,l}^2\cdot \bar{P}_{j},
\end{aligned}
\end{equation}
where
\begin{equation}\label{define of al}
\begin{aligned}
a_{j,l}&=\mathbb{E}_{\mathbf{h}_{j}}\left[\frac{|g_{j,l}|^2}{{\|\mathbf{g}_{j}\|^2}}\right],
\end{aligned}
\end{equation}
with $\sum_{l\in\mathcal{B}}a_{j,l}=1$ for any $j\in\mathcal{K}$. By substituting (\ref{define of w_M}) into (\ref{Rk_mu}), the maximum achievable ergodic rate of user $k$ with MRT, $R_{k}^{M}$, can be further obtained as
\begin{equation}\label{Rk_mu_M}
R_{k}^{M}=\mathbb{E}_{\mathbf{h}_{k}}\left[\log_{2}\left(1+\mu_{k}^M\|\tilde{\mathbf{g}}_{k}\|^{2}\right)\right],
\end{equation}
where the average received SINR $\mu_{k}^{M}$ is given by
\begin{align}\label{mu_M}
\mu_{k}^{M}&=\frac{\bar{P}_{k}\|\bm{\gamma}_{k}\|^{2}}
{N_{0}+\sum_{j\in\mathcal{K}, j\neq k}\sum_{l\in\mathcal{B}}a_{j,l} \cdot \gamma_{k,l}^2\cdot \bar{P}_{j}} \nonumber \\
&\mathop{\approx}\limits^{{P_t}/{N_0} \gg 1}\frac{1}{\sum_{j\in\mathcal{K},j\neq k}\sum_{l\in\mathcal{B}}a_{j,l}\cdot\beta_{k,l}^{2}},
\end{align}
by combining (\ref{power per user}), (\ref{define of beta}-\ref{define of mu}) and (\ref{I_intra}).

\subsection{Maximum Achievable Ergodic Rate $R_{k}^{M}$}
\subsubsection{CA}
With the CA layout, $L$ BS antennas are co-located at the center of the cell. It is shown in Section II that the normalized large-scale fading vector $\bm{\beta}_{k}=\frac{1}{\sqrt{L}}\mathbf{1}_{1\times L}$. The average received SINR $\mu_{k}^{MC}$ is then given by
\begin{equation}\label{mu_MC}
\begin{aligned}
\mu_{k}^{MC}=\frac{L}{K-1},
\end{aligned}
\end{equation}
according to (\ref{mu_M}), and the normalized channel gain $\|\tilde{\mathbf{g}}_{k}\|^{2}{=}\frac{1}{L}\|\mathbf{h}_{k}\|^2$ according to (\ref{nomarlized g}). The maximum achievable ergodic rate of user $k$ with MRT in the CA layout can be therefore obtained from (\ref{Rk_mu_M}) as
\begin{equation}\label{Rk_MC}
R_{k}^{MC}=\int_{0}^{\infty}\frac{x^{L-1}e^{-x}}{(L-1)!}\log_{2}\left(1+\frac{1}{K-1}x\right)dx.
\end{equation}
It can be clearly seen from (\ref{Rk_MC}) that $R_{k}^{MC}$ is independent of user $k$'s position, indicating that all the users achieve the same maximum achievable ergodic rate with the CA layout.

\vspace{2mm}
\subsubsection{DA}
With the DA layout, $L$ BS antennas are uniformly distributed within the cell. In contrast to the CA layout, the average received SINR varies under each realization of the BS antenna topology. Appendix A shows that the average received SINR with the DA layout $\mu_{k}^{MD}$ can be obtained as\footnote{Note that in (\ref{mu_MD}-\ref{Rk_MD}), the normalized large-scale fading coefficients are supposed to be nonidentical. With uniformly distributed BS antennas, $\beta_{k,l}$, which is determined by the positions of user $k$ and BS antenna $l$, is a continuous random variable. The probability that $\beta_{k,l_{1}}=\beta_{k,l_{2}}$ for $l_{1}\neq l_{2}$ is therefore zero.}
\begin{equation}\label{mu_MD}
\begin{aligned}
\mu_{k}^{MD}=\frac{1}{\sum_{j\in\mathcal{K},j\neq k}\sum_{l\in\mathcal{B}}\beta_{k,l}^{2}\sum_{m\in\mathcal{B}, m\neq l}
\frac{\beta_{j,l}^{-2}\beta_{j,m}^{-2}\left(\log\beta_{j,l}^{-2}-\log\beta_{j,m}^{-2}-1\right)+\beta_{j,m}^{-4}}
{\left(\beta_{j,l}^{-2}-\beta_{j,m}^{-2}\right)^{2}}
\prod_{t\in\mathcal{B}, t\neq m, t\neq l}\frac{\beta_{j,t}^{-2}}{\beta_{j,t}^{-2}-\beta_{j,m}^{-2}}}.
\end{aligned}
\end{equation}

Moreover, the normalized channel gain $\|\tilde{\mathbf{g}}_{k}\|^{2}$ is a hypoexponential random variable with the probability density function (pdf)\cite{Ross}
\begin{equation}\label{pdf of NormaliedG}
f_{\|\tilde{\mathbf{g}}_{k}\|^{2}}(x){=}\sum_{l\in\mathcal{B}}\beta_{k,l}^{-2}\exp\left\{-\beta_{k,l}^{-2}x\right\}
\hspace{-0.2cm}\prod_{i\in\mathcal{B}, i\neq l}\frac{\beta_{k,i}^{-2}}{\beta_{k,i}^{-2}{-}\beta_{k,l}^{-2}}.
\end{equation}
By substituting (\ref{pdf of NormaliedG}) into (\ref{Rk_mu_M}), the maximum achievable ergodic rate of user $k$ with MRT in the DA layout can be obtained as
\begin{equation}\label{Rk_MD}
\begin{aligned}
R_{k}^{MD}&{=}\hspace{-0.1cm}\sum_{l\in\mathcal{B}}\exp\hspace{-0.1cm}\left\{\frac{\beta_{k,l}^{-2}}{\mu_{k}^{MD}}\right\}
\hspace{-0.1cm}E_{1}\hspace{-0.1cm}\left\{\frac{\beta_{k,l}^{-2}}{\mu_{k}^{MD}}\right\}\hspace{-0.1cm}
\prod_{i\in\mathcal{B}, i\neq l}\frac{\beta_{k,i}^{-2}}{\beta_{k,i}^{-2}{-}\beta_{k,l}^{-2}}\log_{2}e,
\end{aligned}
\end{equation}
where $E_{1}\left\{x\right\}=\int_{x}^{\infty}t^{-1}e^{-t}dt$.

\begin{figure}[t]
\begin{center}
\includegraphics[width=0.75\textwidth]{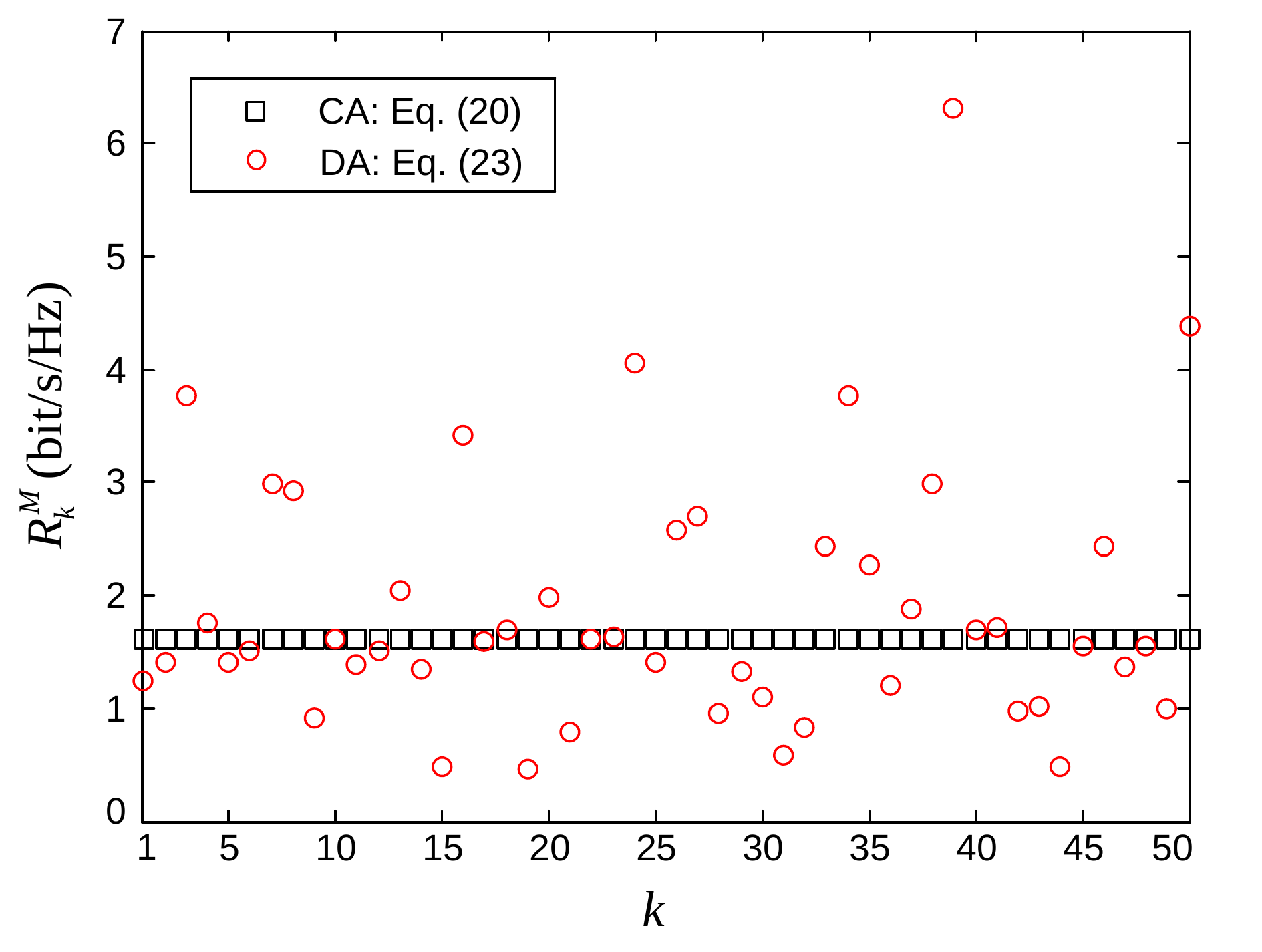}
\caption{Maximum achievable ergodic rate with MRT of each user in the CA layout $R_{k}^{MC}$ and maximum achievable ergodic rate with MRT of each user in the DA layout $R_{k}^{MD}$. The x-axis $k$ denotes the index of a user. $\alpha=4$. $L=100$. $K=50$.}
\label{FIG_Rk_MRT}
\end{center}
\end{figure}

We can see from (\ref{mu_MC}-\ref{Rk_MC}) that with the CA layout, both the average received SINR $\mu_{k}^{MC}$ and the maximum achievable ergodic rate $R_{k}^{MC}$ are solely determined by the number of BS antennas $L$ and the number of users $K$, which are independent of user $k$'s position. With the DA layout, in contrast, (\ref{mu_MD}) and (\ref{Rk_MD}) suggest that the rate performance may significantly vary with users' positions. Fig. \ref{FIG_Rk_MRT} illustrates the maximum achievable ergodic rate $R_{k}^{M}$ of each user with MRT in both the CA layout and the DA layout. Different from the CA case where all the users have the same rate, with the DA layout, the average received SINR of each user is sensitive to its position, thus leading to varying rate performance. We can also observe from Fig. \ref{FIG_Rk_MRT} that despite the fluctuation, a higher rate can be achieved in the DA layout on average. In the next section, we will focus on the average user rate performance and present an asymptotic analysis as the number of BS antennas $L$ and the number of users $K$ grow infinitely while $L/K\rightarrow \upsilon$.

\subsection{Asymptotic Average User Rate $\tilde{R}^{M}$}

\subsubsection{CA}
By combining (\ref{define of AverageR_CA}) and (\ref{Rk_MC}), the average user rate with the CA layout can be easily obtained as
\begin{equation}\label{AverageR_MC}
\bar{R}^{MC}=\int_{0}^{\infty}\frac{x^{L-1}e^{-x}}{(L-1)!}\log_{2}\left(1+\frac{1}{K-1}x\right)dx.
\end{equation}
As $L, K{\rightarrow} \infty$ and $L/K{\rightarrow} \upsilon$, we have
\begin{equation}\label{AsymR_MC}
\begin{aligned}
\tilde{R}^{MC}=\mathop {\lim }\limits_{\scriptstyle L, K\rightarrow \infty,\hfill\atop \scriptstyle L/K\rightarrow \upsilon\hfill}\bar{R}^{MC}
=\log_{2}\left(1+\upsilon\right).
\end{aligned}
\end{equation}

\vspace{2mm}
\subsubsection{DA}
As it is difficult to derive the asymptotic average user rate from (\ref{Rk_MD}), we resort to an upper-bound to study the scaling behavior of the average user rate with the DA layout $\bar{R}^{MD}$ in the following.

Specifically, with a large number of BS antennas $L$, for each user $k\in\mathcal{K}$, there is a high chance that it is very close to some BS antenna $l_{k}^{*}$ such that the large-scale fading coefficient $\gamma_{k,l_{k}^{*}}\gg \gamma_{k,l}$ if $l\neq l_{k}^{*}$. In this case, we have $\beta_{k,l_{k}^{*}}\gg \beta_{k,l}$ and $a_{k,l_{k}^{*}}\gg a_{k,l}$ for $l\neq l_{k}^{*}$ according to (\ref{define of beta}) and (\ref{define of al}), respectively. The maximum achievable ergodic rate of user $k$ with MRT in the DA layout can be then approximately written as
\begin{equation}\label{Rk_MD_approx}
{R_{k}^{MD}}\approx \exp\left\{\frac{1}{{\mu_{k}^{MD}}}\right\}E_{1}\left\{\frac{1}{{\mu_{k}^{MD}}}\right\}\log_{2}e,
\end{equation}
according to (\ref{Rk_MD}), and the average received SINR ${\mu_{k}^{MD}}$ can be approximated from (\ref{mu_M}) as
\begin{align}\label{mu_MD_approx_1}
\mu_{k}^{MD}&\mathop{\approx} \limits^{a_{j,l_{j}^{*}}\gg a_{j,l}, l\neq l_{j}^{*}}
\hspace{-0.5cm}\frac{1}{\sum_{j\in \mathcal{K}, j\neq k}\beta_{k,l_{j}^{*}}^{2}}
=\frac{\|\bm{\gamma}_{k}\|^2}{\sum_{j\in\mathcal{K}, j\neq k}\gamma_{k, l_{j}^{*}}^{2}} \nonumber \\
&\mathop{\approx} \limits^{\gamma_{k,l_{k}^{*}}\gg \gamma_{k,l}, l\neq l_{k}^{*}} \hspace{-0.5cm}\frac{\gamma_{k,l_{k}^{*}}^{2}}{\sum_{j\in\mathcal{K}, j\neq k, l_{j}^{*}=l_{k}^{*}}\gamma_{k,l_{j}^{*}}^{2}{+}\sum_{j\in\mathcal{K}, j\neq k, l_{j}^{*}\neq l_{k}^{*}}\gamma_{k,l_{j}^{*}}^{2}} \nonumber \\
&=\frac{1}{m_{k}+\sum_{j\in\mathcal{K}, j\neq k, l_{j}^{*}\neq l_{k}^{*}}\frac{\gamma_{k,l_{j}^{*}}^{2}}{\gamma_{k,l_{k}^{*}}^{2}}},
\end{align}
where $m_{k}=|\mathcal{K}_{k}|$, with $\mathcal{K}_{k}$ denoting the set of users whose closest BS antenna is the same as user $k$'s, i.e., $j\in\mathcal{K}_{k}$ if and only if $l_{j}^{*}=l_{k}^{*}$ for $j\neq k$ and $j\in \mathcal{K}$. With a large number of BS antennas $L$, the access distance from each user $j\in\mathcal{K}$ to its closest BS antenna $l_{j}^{*}$ is very small such that the large-scale fading coefficient $\gamma_{k,l_{j}^{*}}\approx \gamma_{k, j}$. The average received SINR $\mu_{k}^{MD}$ given in (\ref{mu_MD_approx_1}) can be then written as
\begin{align}\label{mu_MD_approx_2}
\mu_{k}^{MD}\approx \frac{1}{m_{k}+\sum_{j\in\mathcal{K}, j\neq k, l_{j}^{*}\neq l_{k}^{*}}\frac{\gamma_{k,j}^{2}}{\gamma_{k,l_{k}^{*}}^{2}}},
\end{align}
which is upper-bounded by
\begin{align}\label{mu_MD_ub}
\mu_{k}^{MD}\leq \mu _k^{{MD}-ub} =\left\{ {\begin{array}{*{20}{c}}
{\frac{1}{{{m_k}}}}&{\text{if}\quad{m_k} \ne 0,}\\
{{\left(\tfrac{d_{k}^{j_{(1)}}}{d_{k}^{l_{(1)}}}\right)^\alpha }}&{\text{otherwise},}
\end{array}} \right.
\end{align}
where $d_{k}^{l_{(1)}}$ and $d_{k}^{j_{(1)}}$ denote the minimum access distances from user $k$ to $L$ BS antennas and the other $K-1$ users, respectively. By combining (\ref{Rk_MD_approx}) and (\ref{mu_MD_ub}), an upper-bound of the average user rate $\bar{R}^{MD}$ can be obtained as
\begin{align}\label{AverageR_MD_ub}
{\bar{R}^{MD}}&{\leq}{\bar{R}^{MD-ub}} \nonumber \\
&{=}\mathbb{E}_{\left\{\mathbf{r}_{k}^{U}\right\}_{k\in\mathcal{K}}, \left\{\mathbf{r}_{l}^{B}\right\}_{l\in\mathcal{B}}}
\hspace{-0.1cm}\left\{\exp\hspace{-0.1cm}\left(\hspace{-0.1cm}\tfrac{1}{{\mu_{k}^{MD-ub}}}\hspace{-0.1cm}\right)\hspace{-0.1cm}
E_{1}\hspace{-0.1cm}\left(\hspace{-0.1cm}\tfrac{1}{{\mu_{k}^{MD-ub}}}\hspace{-0.1cm}\right)\hspace{-0.1cm}\log_{2}e\right\}.
\end{align}
As $L, K\rightarrow \infty$ and $L/K\rightarrow \upsilon$, Appendix B shows that the asymptotic upper-bound is given by
\begin{align}\label{AsymR_MD_ub}
\tilde{R}^{MD-ub}&{=}\hspace{-0.3cm}\mathop{\lim} \limits_{\scriptstyle L, K\rightarrow \infty,\hfill\atop \scriptstyle L/K\rightarrow \upsilon\hfill}\hspace{-0.2cm}\bar{R}^{MD-ub}
{=}\left(\sum_{n=1}^{\infty}\exp\left\{n\right\}\hspace{-0.1cm}E_{1}\left\{n\right\}\hspace{-0.1cm} \frac{\upsilon^{-n}e^{-\frac{1}{\upsilon}}}{n!}{+} \right. \nonumber \\
&\left. \hspace{-0.1cm}\int_{0}^{\infty}\hspace{-0.3cm}\exp\hspace{-0.1cm}\left\{\hspace{-0.1cm}y^{-\alpha}{-}\frac{1}{\upsilon}\right\}\hspace{-0.1cm}
E_{1}\hspace{-0.1cm}\left\{y^{-\alpha}\right\}\hspace{-0.1cm} \frac{2\upsilon y}
{(\upsilon{+}y^2)^2}dy\hspace{-0.1cm}\right)\hspace{-0.1cm}\log_{2}e.
\end{align}

\begin{figure}[t]
\begin{center}
\includegraphics[width=0.75\textwidth]{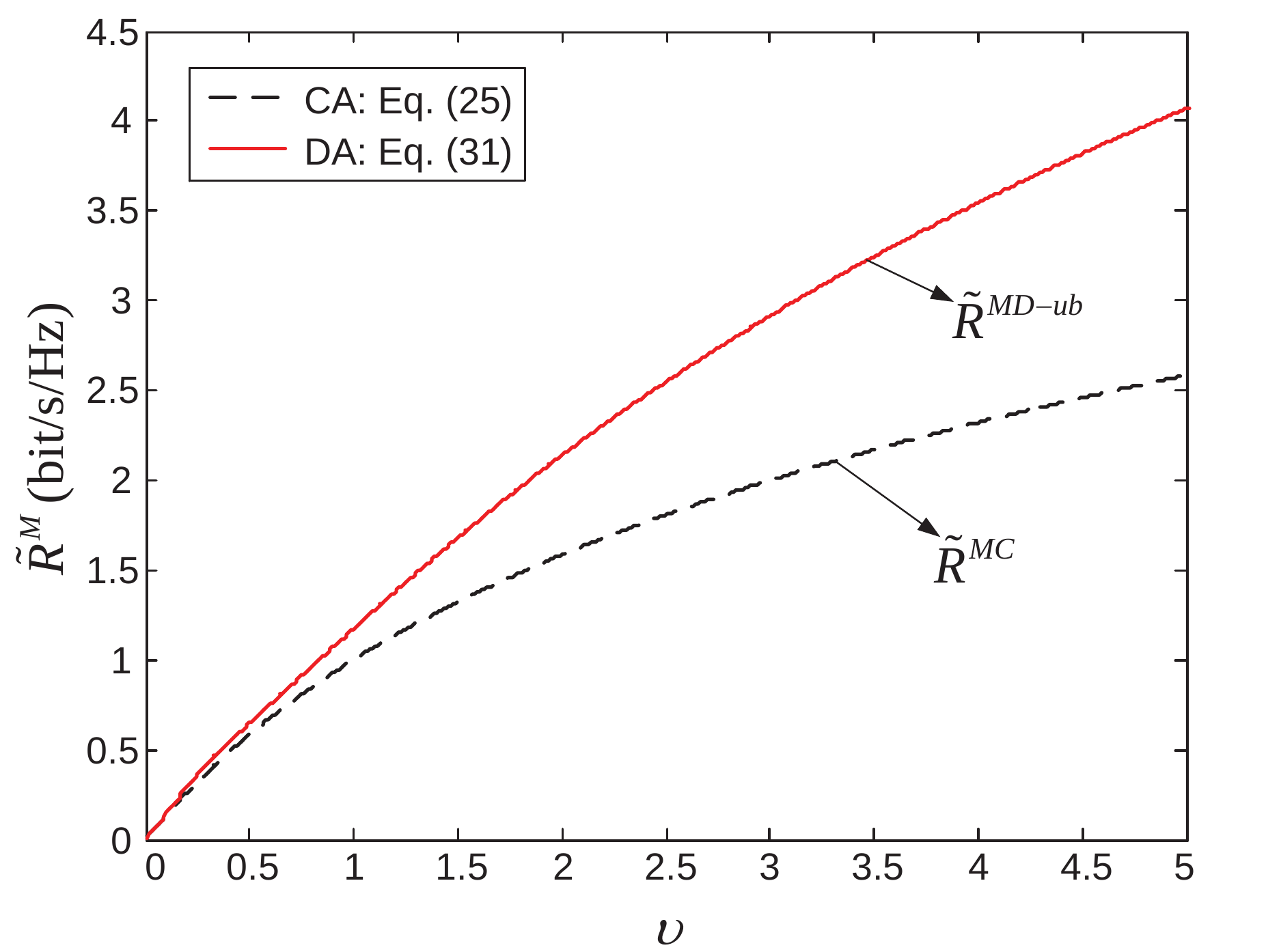}
\caption{Asymptotic average user rate with MRT in the CA layout $\tilde{R}^{MC}$ and asymptotic upper-bound of the average user rate with MRT in the DA layout $\tilde{R}^{MD-ub}$ versus the ratio $\upsilon$ of the number of BS antennas $L$ and the number of users $K$. $\alpha=4$.}
\label{FIG_Rk_M_V}
\end{center}
\end{figure}

Fig. \ref{FIG_Rk_M_V} presents the asymptotic average user rate with the CA layout, $\tilde{R}^{MC}$, and the asymptotic upper-bound of the average user rate with the DA layout, $\tilde{R}^{MD-ub}$. As we can observe from Fig. \ref{FIG_Rk_M_V}, although both $\tilde{R}^{MC}$ and $\tilde{R}^{MD-ub}$ logarithmically increase with $\upsilon$, a much higher rate is achieved in the DA case when $\upsilon$ is large. In particular, a large $\upsilon$ indicates that the number of BS antennas $L$ is much higher than the number of users $K$. In this case, for any user $k$, the number of interfering users whose closest antenna is the same as user $k$'s is approximately zero, i.e., $m_{k}\approx 0$. Moreover, the minimum access distances $d_{k}^{l_{(1)}}$ and $d_{k}^{j_{(1)}}$ decrease in the orders of $1/{\sqrt{L}}$ and $1/{\sqrt{K-1}}$, respectively, as the number of BS antennas $L$ and the number of users $K$ increase. As a result, we can see from (\ref{mu_MD_ub}) that the upper-bound of the average received SINR $\mu_{k}^{MD-ub}$ scales in the order of $\upsilon^{\frac{\alpha}{2}}$ when $L, K{\rightarrow} \infty$ and $L/K{\rightarrow} \upsilon$. We can then conclude from (\ref{AverageR_MD_ub}-\ref{AsymR_MD_ub}) that the asymptotic upper-bound $\tilde{R}^{MD-ub}$ increases with $\upsilon$ in the order of $\frac{\alpha}{2}\log_{2}\upsilon$, which is higher than  $\tilde{R}^{MC}$ according to (\ref{AsymR_MC}), as the path-loss factor $\alpha>2$. The gap between $\tilde{R}^{MD-ub}$ and $\tilde{R}^{MC}$ is further enlarged as $\upsilon$ increases.

\subsection{Simulation Results}
In this section, simulation results are presented to verify the above average user rate analysis with MRT. As described in Section II, $K$ users are supposed to be uniformly distributed in a circular cell with radius 1. With the CA layout, all $L$ BS antennas are co-located at the center of the cell, and simulation results are obtained by averaging over 500 realizations of users' positions. With the DA layout, $L$ BS antennas are uniformly distributed within the cell, and the simulation results are further averaged over 50 realizations of BS antennas' positions.


\begin{figure}[t]
\begin{center}
\includegraphics[width=0.75\textwidth]{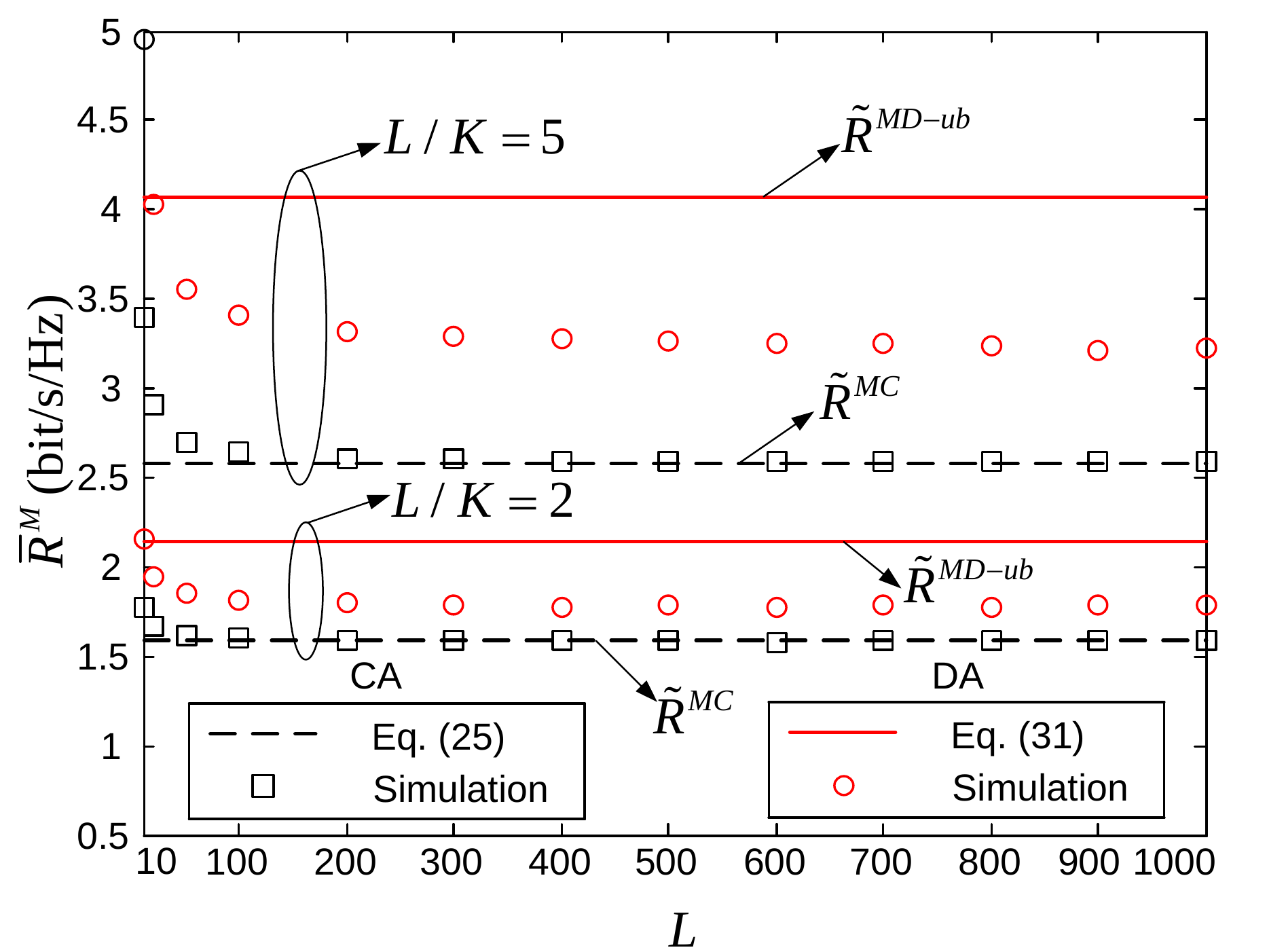}
\caption{Average user rate with MRT $\bar{R}^{M}$ versus the number of BS antennas $L$ in the CA layout and the DA layout. $P_{t}/N_{0}=20$dB. $\alpha=4$. $L/K=2, 5$.}
\label{FIG_Rk_M_FixedV}
\end{center}
\end{figure}

Fig. \ref{FIG_Rk_M_FixedV} presents the simulation results of the average user rate with MRT in both the CA layout and the DA layout under different values of the number of BS antennas $L$ and the number of users $K$ with the ratio of $L$ and $K$ fixed. With the CA layout, as $L, K{\rightarrow} \infty$ and $L/K{\rightarrow} \upsilon$, the asymptotic average user rate $\tilde{R}^{MC}$ has been derived in (\ref{AsymR_MC}) as a function of $\upsilon$, and is plotted in Fig. \ref{FIG_Rk_M_FixedV}. As we can see from this figure, with a large number of BS antennas $L$ and a large number of users $K$, the asymptotic rate $\tilde{R}^{MC}$ serves as a good approximation for the average user rate $\bar{R}^{MC}$.

With the DA layout, an asymptotic upper-bound of the average user rate $\tilde{R}^{MD-ub}$ is derived in (\ref{AsymR_MD_ub}). We can clearly observe from Fig. \ref{FIG_Rk_M_FixedV} that similar to the asymptotic upper-bound $\tilde{R}^{MD-ub}$, as the number of BS antennas $L$ and the number of users $K$ grow with a fixed ratio, the average user rate with the DA layout $\bar{R}^{MD}$ approaches a constant that is solely determined by the ratio of $L$ and $K$. It is always higher than the average user rate with the CA layout $\bar{R}^{MC}$, and the gap is enlarged as the ratio of $L$ and $K$ increases.

\section{Rate Analysis with ZFBF}
With MRT, the intra-cell interference is a severe limiting factor for the rate performance no matter which BS antenna layout is adopted. In this section, we focus on an orthogonal linear precoding scheme, ZFBF \cite{Caire}, with which the precoding vector $\mathbf{w}_{k}$ of each user is selected such that $\tilde{\mathbf{g}}_{j}\mathbf{w}_{k}=0$ for $j\neq k, j\in\mathcal{K}$. Specifically, let $\tilde{\mathbf{G}}=[\tilde{\mathbf{g}}_{1}^{T}, \tilde{\mathbf{g}}_{2}^{T}, ..., \tilde{\mathbf{g}}_{K}^{T}]^{T}$, and $\mathbf{F}$ denotes the pseudo-inverse of $\tilde{\mathbf{G}}$, i.e, $\mathbf{F}=\tilde{\mathbf{G}}^{\dag}(\tilde{\mathbf{G}}\tilde{\mathbf{G}}^{\dag})^{-1}$. The precoding vector $\mathbf{w}_{k}$ can be then written as
\begin{equation}\label{define of w_Z}
\mathbf{w}_{k}=\frac{\mathbf{f}_{k}}{\|\mathbf{f}_{k}\|},
\end{equation}
$k\in\mathcal{K}$, where $\mathbf{f}_{k}$ is the $k$th column vector of $\mathbf{F}$. By combining (\ref{Rk_mu}) and (\ref{define of w_Z}), the maximum achievable ergodic rate of user $k$ with ZFBF can be written as
\begin{equation}\label{Rk_mu_Z}
R_{k}^{Z}=\mathbb{E}_{\mathbf{h}_{k}}\left[\log_{2}\left(1+\frac{\mu_{k}^{Z}}{\|\mathbf{f}_{k}\|^{2}}\right)\right],
\end{equation}
where the average received signal-to-noise ratio (SNR) $\mu_{k}^{Z}$ is given by
\begin{equation}\label{mu_Z}
\mu_{k}^{Z}=\frac{\bar{P}_{k}}{N_{0}}\|\bm{\gamma}_{k}\|^2,
\end{equation}
according to (\ref{define of mu}), as there is no intra-cell interference. Let us define $1/\|\mathbf{f}_{k}\|^{2}$ as the effective channel gain of user $k$, which can be further written as \cite{Matrix_Analysis, Gore}
\begin{equation}\label{EffectiveChannelGain}
\frac{1}{\|\mathbf{f}_{k}\|^{2}}=\frac{1}{[(\tilde{\mathbf{G}}\tilde{\mathbf{G}}^{\dag})^{-1}]_{k,k}}
\mathop{=} \limits^{\mathbf{Z}=\tilde{\mathbf{G}}\tilde{\mathbf{G}}^{\dag}}\frac{\det\mathbf{Z}}{\det\mathbf{Z}_{k}}=\mathbf{Z}_{k}^{sc},
\end{equation}
where $\mathbf{Z}_{k}$ denotes the submatrix of $\mathbf{Z}$ by deleting the $k$th row and the $k$th column, and $\mathbf{Z}_{k}^{sc}$ denotes the Schur complement of $\mathbf{Z}_{k}$. We can clearly see from (\ref{Rk_mu_Z}) that with ZFBF, the rate performance is closely dependent on the average received SNR and the effective channel gain, both of which are determined by the positions of BS antennas and users.

Note that to perform ZFBF, the number of BS antennas $L$ should be no smaller than the number of users $K$. In the following subsections, a special focus will be put on the case with $L\gg K$.

\subsection{Maximum Achievable Ergodic Rate $R_{k}^{Z}$}

\subsubsection{CA}
With the CA layout, all the BS antennas are co-located at the center of the cell, i.e., $\mathbf{r}_{l}^{B}=(0, 0)$ for any $l\in\mathcal{B}$. For user $k$ located at $\mathbf{r}_{k}^{U}=(\rho_{k}, \theta_{k})$, it can be easily obtained from (\ref{define of gamma}) that the large-scale fading vector $\bm{\gamma}_{k}=\rho_{k}^{-\alpha}\mathbf{1}_{1\times L}$. The average received SNR $\mu_{k}^{ZC}$ can be then obtained as
\begin{equation}\label{mu_ZC}
\mu_{k}^{ZC}=\frac{P_{t}L}{KN_{0}}\rho_{k}^{-\alpha},
\end{equation}
according to (\ref{power per user}) and (\ref{mu_Z}). As the normalized large-scale fading vector $\bm{\beta}_{k}=\frac{1}{\sqrt{L}}\mathbf{1}_{1\times L}$, we have $\mathbf{Z}=\tilde{\mathbf{G}}\tilde{\mathbf{G}}^{\dag}=\frac{1}{L}\mathbf{H}\mathbf{H}^{\dag}\sim \mathcal{W}_{K}(L, \frac{1}{L}\mathbf{I}_{K})$, where $\mathbf{H}=[\mathbf{h}_{1}^{T}, \mathbf{h}_{2}^{T}, ..., \mathbf{h}_{K}^{T}]^{T}$. According to Theorem 3. 2. 10 in \cite{Multivariate}, the effective channel gain $1/\|\mathbf{f}_{k}\|^2=\mathbf{Z}_{k}^{sc}\sim \mathcal{W}_{1}(L-K+1, \frac{1}{L})$. 
The maximum achievable ergodic rate of user $k$ with ZFBF in the CA layout can be then obtained from (\ref{Rk_mu_Z}) and (\ref{mu_ZC}) as
\begin{equation}\label{Rk_ZC}
R_{k}^{ZC}{=}\hspace{-0.1cm}\int_{0}^{\infty}\hspace{-0.1cm}\frac{L(Lx)^{L-K}e^{-Lx}}{(L-K)!}\log_{2}\left(1{+}\frac{P_{t}L\rho_{k}^{-\alpha}}{KN_{0}}x\right)dx.
\end{equation}
Note that as $L/K$ grows, the effective channel gain $1/{\|\mathbf{f}_{k}\|^2}$ becomes increasingly deterministic, and eventually converges to $\frac{L-K+1}{L}$. We then have
\begin{align}\label{Rk_ZC_approx}
R_{k}^{ZC}&\mathop{\approx} \limits^{L\gg K}\log_{2}\left(1+\frac{P_{t}(L-K+1)}{KN_{0}}\rho_{k}^{-\alpha}\right) \nonumber \\
&\mathop{\approx} \limits^{K\gg 1}\log_{2}\left(1+\frac{P_{t}}{N_{0}}\left(\frac{L}{K}-1\right)\rho_{k}^{-\alpha}\right)\nonumber \\
&\mathop{\approx} \limits^{P_{t}/N_{0}\gg 1} \log_{2}\left(\frac{P_{t}}{N_{0}}\left(\frac{L}{K}-1\right)\rho_{k}^{-\alpha}\right).
\end{align}
In contrast to (\ref{Rk_MC}), $R_{k}^{ZC}$ depends on the access distance $\rho_{k}$, which indicates that with ZFBF, users far away from the cell center suffer from significant degradation of rate performance. 

\vspace{2mm}
\subsubsection{DA}
With the DA layout, it is difficult to derive the distribution of the effective channel gain $1/\|\mathbf{f}_{k}\|^2$. Therefore, we resort to a lower-bound to study the scaling behavior of the rate performance. Appendix C shows that with $L\gg K$, the maximum achievable ergodic rate of user $k$  with ZFBF in the DA layout $R_{k}^{ZD}$ is lower-bounded by
\begin{align}\label{Rk_ZD_lb}
R_{k}^{ZD}&{\geq} R_{k}^{ZD-lb} \nonumber \\
&{=}\exp\left\{\hspace{-0.1cm}\tfrac{KN_{0}}{P_{t}}\hspace{-0.1cm}\left(\tilde{d}_{k}^{l_{(1)}}\hspace{-0.1cm}\right)^{\alpha}\right\}
\hspace{-0.1cm}E_{1}\hspace{-0.1cm}\left\{\hspace{-0.1cm}\tfrac{KN_{0}}{P_{t}}\hspace{-0.1cm}\left(\tilde{d}_{k}^{l_{(1)}}\hspace{-0.1cm}\right)^{\alpha}\right\}\hspace{-0.1cm}\log_{2}e,
\end{align}
where $\tilde{d}_{k}^{l_{(1)}}$ denotes the minimum access distance from user $k$ to $L-K+1$ uniformly distributed BS antennas.

\begin{figure}[t]
\begin{center}
\includegraphics[width=0.75\textwidth]{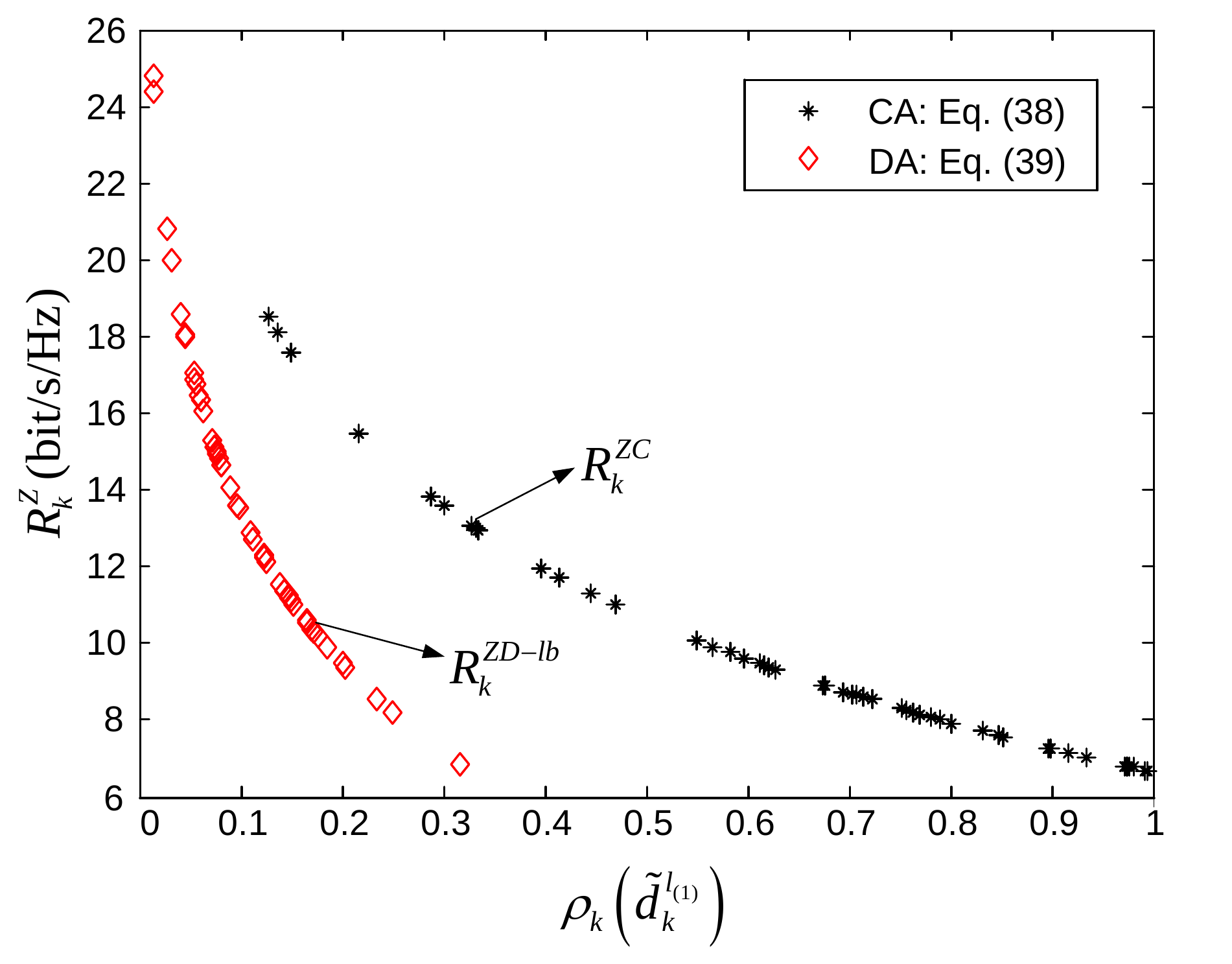}
\caption{Maximum achievable ergodic rate with ZFBF of each user in the CA layout $R_{k}^{ZC}$ and the lower-bound of maximum achievable ergodic rate with ZFBF of each user in the DA layout $R_{k}^{ZD-lb}$. $P_{t}/N_{0}=20$dB. $\alpha=4$. $L=100$. $K=50$.}
\label{FIG_Rk_ZFBF}
\end{center}
\end{figure}

We can see from (\ref{Rk_ZC_approx}) and (\ref{Rk_ZD_lb}) that both $R_{k}^{ZC}$ and $R_{k}^{ZD-lb}$ are crucially determined by the minimum access distance of user $k$.\footnote{With the CA layout, the minimum access distance of user $k$ is equal to its radial coordinate $\rho_{k}$ as all the BS antennas are co-located at the center of the cell.} Fig. \ref{FIG_Rk_ZFBF} presents the maximum achievable ergodic rate with ZFBF of each user in the CA layout $R_{k}^{ZC}$ and the lower-bound of the maximum achievable ergodic rate with ZFBF of each user in the DA layout $R_{k}^{ZD-lb}$. We can observe from Fig. \ref{FIG_Rk_ZFBF} that although for given minimum access distance, $R_{k}^{ZD-lb}$ is always lower than $R_{k}^{ZC}$, the chance that each user has a small minimum access distance in the DA layout is much higher than that in the CA case. We can then expect that on average, a higher rate can be achieved in the DA layout. In the next section, we will focus on the average user rate performance and present an asymptotic analysis as the number of BS antennas $L$ and the number of users $K$ grow infinitely while $L/K\rightarrow \upsilon >1$.

\subsection{Asymptotic Average User Rate $\tilde{R}^{Z}$}

\subsubsection{CA}
By combining (\ref{define of AverageR_CA}) and (\ref{Rk_ZC_approx}), the average user rate with the CA layout can be obtained as
\begin{align}\label{AverageR_ZC}
\bar{R}^{ZC}&=\int_{0}^{1}\log_{2}\left(\frac{P_{t}}{N_{0}}\left(\frac{L}{K}-1\right)x^{-\alpha}\right)\cdot 2xdx \nonumber \\
&=\log_{2}\left(\frac{P_{t}}{N_{0}}\left(\frac{L}{K}-1\right)\right)+\frac{\alpha}{2}\log_{2}e.
\end{align}
It is clear from (\ref{AverageR_ZC}) that as $L, K\rightarrow \infty$ and $L/K\rightarrow \upsilon > 1$, the asymptotic average user rate is given by
\begin{equation}\label{AsymR_ZC}
\tilde{R}^{ZC}{=}\hspace{-0.1cm}\mathop{\lim}\limits_{\scriptstyle L, K\rightarrow \infty \hfill \atop \scriptstyle L/K\rightarrow \upsilon \hfill}\hspace{-0.1cm}\bar{R}^{ZC}
{=}\log_{2}\left(\frac{P_{t}}{N_{0}}(\upsilon-1)\right){+}\frac{\alpha}{2}\log_{2}e.
\end{equation}
We can see from (\ref{AsymR_MC}) and (\ref{AsymR_ZC}) that similar to the MRT case, with the CA layout, the asymptotic average user rate with ZFBF is also determined by the ratio $\upsilon$ of the number of BS antennas $L$ and the number of users $K$. By eliminating the intra-cell interference through joint precoding among users, however, a much higher rate can be achieved with ZFBF especially when $P_{t}/N_{0}$ or $\upsilon$ is large.

\vspace{2mm}
\subsubsection{DA}
According to (\ref{Rk_ZD_lb}), a lower-bound of the average user rate with ZFBF in the DA layout can be written as
\begin{align}\label{AverageR_ZD_lb_1}
\bar{R}^{ZD-lb}&{=}\mathbb{E}_{\left\{\mathbf{r}_{k}^{U}\right\}_{k\in\mathcal{K}},\left\{\mathbf{r}_{l}^{B}\right\}_{l\in\mathcal{B}}}
\left[R_{k}^{ZD-lb}\right] \nonumber \\
&{=}\mathbb{E}_{\tilde{d}_{k}^{l_{(1)}}}\hspace{-0.2cm}\left[\exp\hspace{-0.1cm}\left\{\hspace{-0.1cm}\tfrac{KN_{0}}{P_{t}}\hspace{-0.1cm}\left(\tilde{d}_{k}^{l_{(1)}}\hspace{-0.1cm}\right)^{\alpha}\right\}
\hspace{-0.1cm}E_{1}\hspace{-0.1cm}\left\{\hspace{-0.1cm}\tfrac{KN_{0}}{P_{t}}\hspace{-0.1cm}\left(\tilde{d}_{k}^{l_{(1)}}\hspace{-0.1cm}\right)^{\alpha}\right\}\hspace{-0.1cm}\log_{2}e\right],
\end{align}
where $\tilde{d}_{k}^{l_{(1)}}$ is the minimum access distance from user $k$ to $L-K+1$ uniformly distributed BS antennas. It has been shown in \cite{Dai_JSAC} that with $L-K+1$ BS antennas uniformly distributed in a disk with radius 1, the minimum access distance $\tilde{d}_{k}^{l_{(1)}}$ given user $k$'s position at $(\rho_{k}, \theta_{k})$ follows the pdf
\begin{equation}\label{pdf of dmin'}
f_{\tilde{d}_{k}^{l_{(1)}}|\rho_{k}}(x|y)=(L-K+1)(1-F(x;y))^{L-K}f(x;y),
\end{equation}
where
\begin{equation}\label{pdf of dl}
\begin{aligned}
f(x; y)&{=}\left\{ {\begin{array}{*{20}{c}}{2x}&{0 {\le} x {\le} 1 {-} y},\\
{\tfrac{{2x}}{\pi }\arccos \tfrac{{{x^2} {+} {y^2} {-} 1}}{{2yx}}}&{1 {-} y {<} x{\le} 1 {+} y},
\end{array}} \right.
\end{aligned}
\end{equation}
and
\begin{equation}\label{cdf of dl}
\begin{aligned}
F(x; y)&{=}\hspace{-0.1cm}\left\{ {\begin{array}{*{20}{c}}
{{x^2}}&{\hspace{-0.6cm}0 {\le} x {\le} 1{ -} y},\\
\begin{array}{l}
\hspace{-0.4cm}\tfrac{{{x^2}}}{\pi }\arccos \tfrac{{{x^2} {+} {y^2} {-} 1}}{{2yx}}{+} \\
\hspace{-0.4cm}\tfrac{1}{\pi }\arccos \tfrac{{1 {+} {y^2} {-} {x^2}}}{{2y}} {-} \tfrac{2}{\pi }{S_\Delta }
\end{array}&{\hspace{-0.6cm}1 {-} y {<} x {\le} 1 {+} y},
\end{array}} \right.
\end{aligned}
\end{equation}
with
\begin{equation}\label{s_delta}
S_{\Delta}{=}\sqrt{{\tfrac{x+y+1}{2}}\left({\tfrac{x+y+1}{2}}{-}1\right)\left({\tfrac{x+y+1}{2}}
{-}x\right)\left({\tfrac{x+y+1}{2}}{-}y\right)}.
\end{equation}
Moreover, with users uniformly distributed in a disk with radius 1, the radial coordinate $\rho_{k}$ follows the pdf $f_{\rho_{k}}(y)=2y$. The lower-bound of the average user rate $\bar{R}^{ZD-lb}$ can be then obtained as
\begin{align}\label{AverageR_ZD_lb}
\bar{R}^{ZD-lb}&{=}2(L-K+1)\log_{2}e\int_{0}^{1}y\int_{0}^{1+y}\exp\left\{\tfrac{KN_{0}}{P_{t}}x^{\alpha}\right\} \nonumber \\
&E_{1}\left\{\tfrac{KN_{0}}{P_{t}}x^{\alpha}\right\} (1-F(x;y))^{L-K}f(x;y)dxdy.
\end{align}
Appendix D further shows that as $L, K{\rightarrow} \infty$ and $L/K{\rightarrow} \upsilon {>} 1$, the asymptotic lower-bound of the average user rate $\tilde{R}^{ZD-lb}$ is given by
\begin{equation}\label{AsymR_ZD}
\tilde{R}^{ZD-lb}=\mathop{\lim} \limits_{\scriptstyle L, K\rightarrow \infty, \hfill \atop \scriptstyle L/K\rightarrow \upsilon}\bar{R}^{ZD-lb}=\infty.
\end{equation}
Recall that it has been shown in (\ref{AsymR_ZC}) that the asymptotic average user rate with the CA layout logarithmically increases with the ratio $\upsilon$ as $L,K\to \infty$ and $L/K\to \upsilon >1$. Here we can see that with the DA layout, the average user rate grows with $L$ unboundedly, which implies substantial rate gains over the CA layout when the number of BS antennas $L$ is large.

\begin{figure}[t]
\begin{center}
\includegraphics[width=0.75\textwidth]{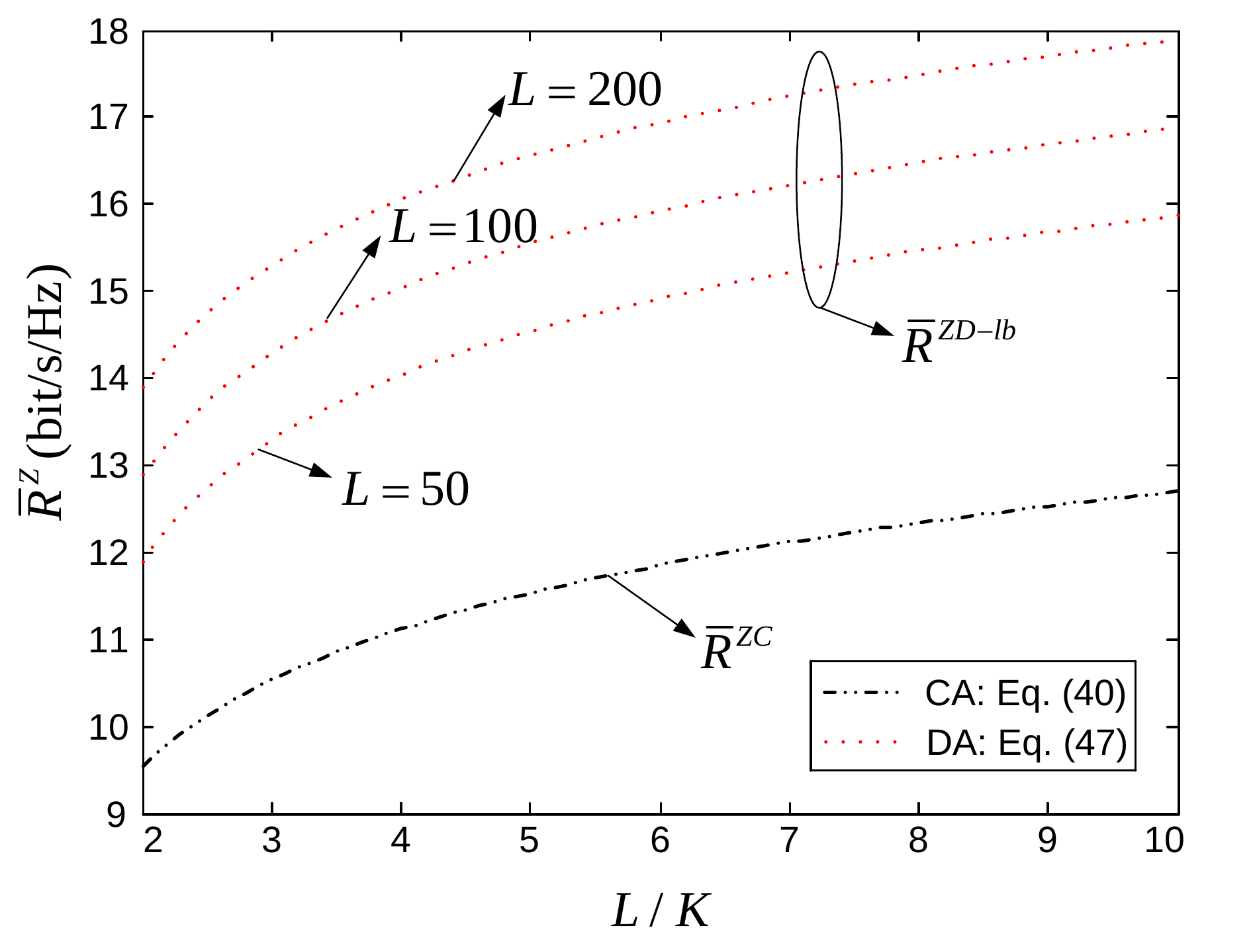}
\caption{Average user rate with ZFBF in the CA layout $\bar{R}^{ZC}$ and the lower-bound of the average user rate with ZFBF in the DA layout $\bar{R}^{ZD-lb}$ versus the ratio of the number of BS antennas $L$ and the number of users $K$. $P_{t}/N_{0}=20$dB. $\alpha=4$.}
\label{FIG_Rk_Z_V}
\end{center}
\end{figure}

To take a closer look at the scaling behavior of the average user rate in the DA layout, Fig. \ref{FIG_Rk_Z_V} plots $\bar{R}^{ZD-lb}$ under various values of the number of BS antennas $L$ and the number of users $K$. As we can see from Fig. \ref{FIG_Rk_Z_V}, similar to $\bar{R}^{ZC}$, $\bar{R}^{ZD-lb}$ also logarithmically increases with the ratio of the number of BS antennas $L$ and the number of users $K$. Nevertheless, in contrast to the CA case where $\bar{R}^{ZC}$ is solely determined by $L/K$, $\bar{R}^{ZD-lb}$ further depends on the number of BS antennas $L$. In fact, (\ref{Rk_ZD_lb}) has shown that with the DA layout, the lower-bound $R^{ZD-lb}$ is determined by the minimum access distance $\tilde{d}_{k}^{l_{(1)}}$, which decreases in the order of $1/\sqrt{L-K+1}$ as $L$ increases according to (\ref{pdf of dmin'}). As a result, $\bar{R}^{ZD-lb}$ scales in the order of $\log_{2}\left(\left(L-K+1\right)^{\frac{\alpha}{2}}/K\right)$. As $L, K{\rightarrow} \infty$ and $L/K{\rightarrow} \upsilon {>} 1$, $\bar{R}^{ZD-lb}$ grows unboundedly, which is in sharp contrast to the CA case where the asymptotic average user rate is a logarithmic function of $\upsilon$. As we can observe from Fig. \ref{FIG_Rk_Z_V}, for given ratio of $L$ and $K$, a much higher rate can be always achieved in the DA layout, and the gap between $\bar{R}^{ZD-lb}$ and $\bar{R}^{ZC}$ is greatly enlarged as $L$ increases.

\subsection{Simulation Results}
\begin{figure}[t]
\begin{center}
\includegraphics[width=0.75\textwidth]{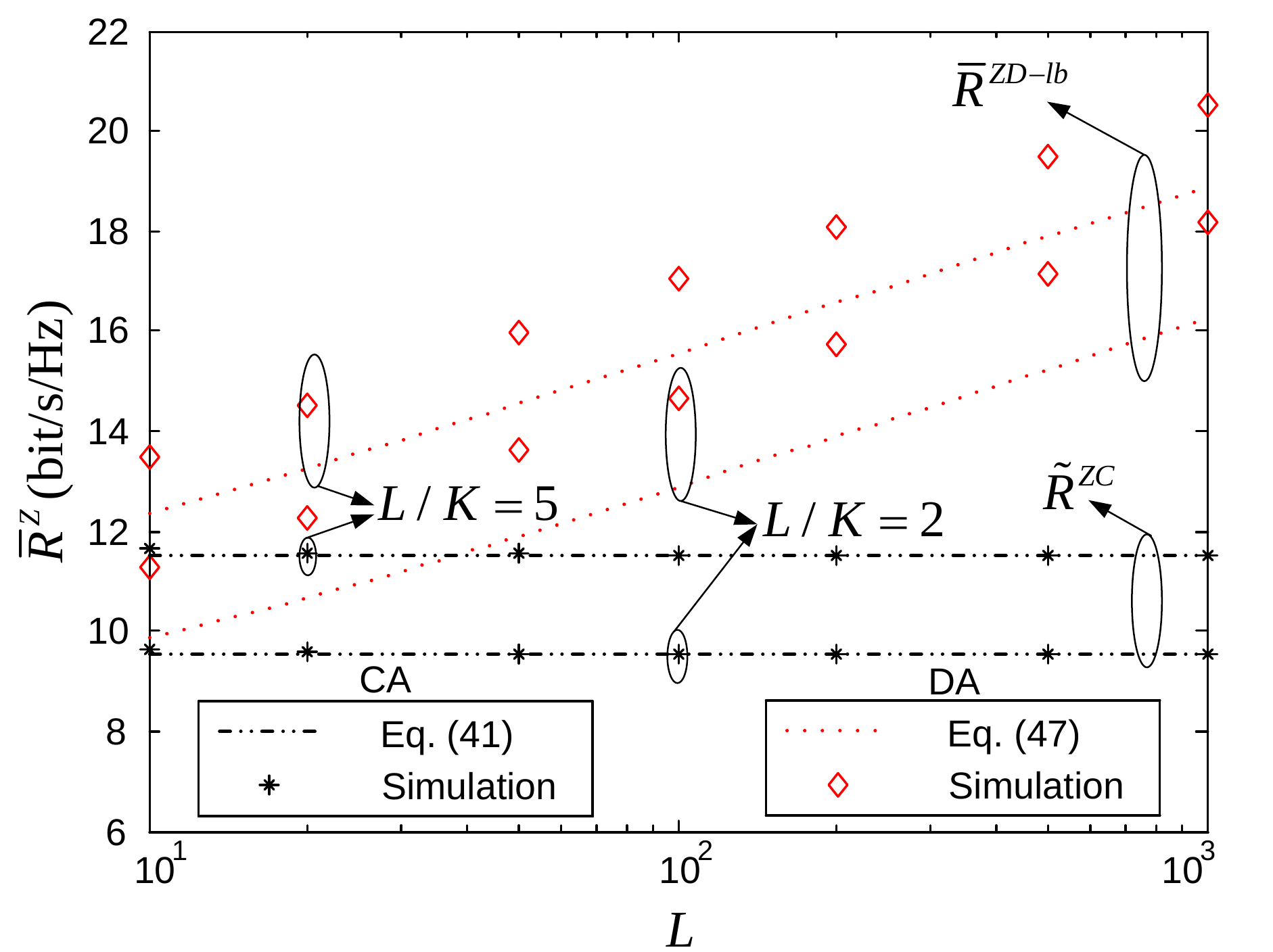}
\caption{Average user rate with ZFBF $\bar{R}^{Z}$ versus the number of BS antennas $L$ in the CA layout and the DA layout. $P_{t}/N_{0}=20$dB. $\alpha=4$. $L/K=2, 5.$}
\label{FIG_Rk_Z_FixedV}
\end{center}
\end{figure}
Simulation results in this section are presented to verify the above average user rate analysis with ZFBF. The simulation setting is the same as that presented in Section III-C. Fig. \ref{FIG_Rk_Z_FixedV} presents the simulation results of the average user rate with ZFBF in the CA layout and the DA layout. With the CA layout,  the asymptotic average user rate $\tilde{R}^{ZC}$ as $L, K{\rightarrow} \infty$ and $L/K{\rightarrow} \upsilon {>} 1$ has been derived as a function of $\upsilon$ in (\ref{AsymR_ZC}), which, as we can see from Fig. \ref{FIG_Rk_Z_FixedV}, serves as a good approximation for the average user rate $\bar{R}^{ZC}$ even when the number of BS antennas $L$ and the number of users $K$ are small.

With the DA layout, a lower-bound of the average user rate $\bar{R}^{ZD-lb}$ is developed in (\ref{AverageR_ZD_lb}), which is shown to be increasing with the number of BS antennas $L$ unboundedly as $L, K{\rightarrow} \infty$ and $L/K{\rightarrow} \upsilon {>}1$. As we can see from Fig. \ref{FIG_Rk_Z_FixedV}, in contrast to the CA case where the average user rate $\bar{R}^{ZC}$ is solely determined by the ratio of $L$ and $K$, the average user rate in the DA layout $\bar{R}^{ZD}$ is, similar to its lower-bound $\bar{R}^{ZD-lb}$, significantly improved as $L$ grows. With a large $L$, substantial rate gains can be achieved in the DA layout.

\begin{figure}[!t]
\begin{center}
\includegraphics[width=0.75\textwidth]{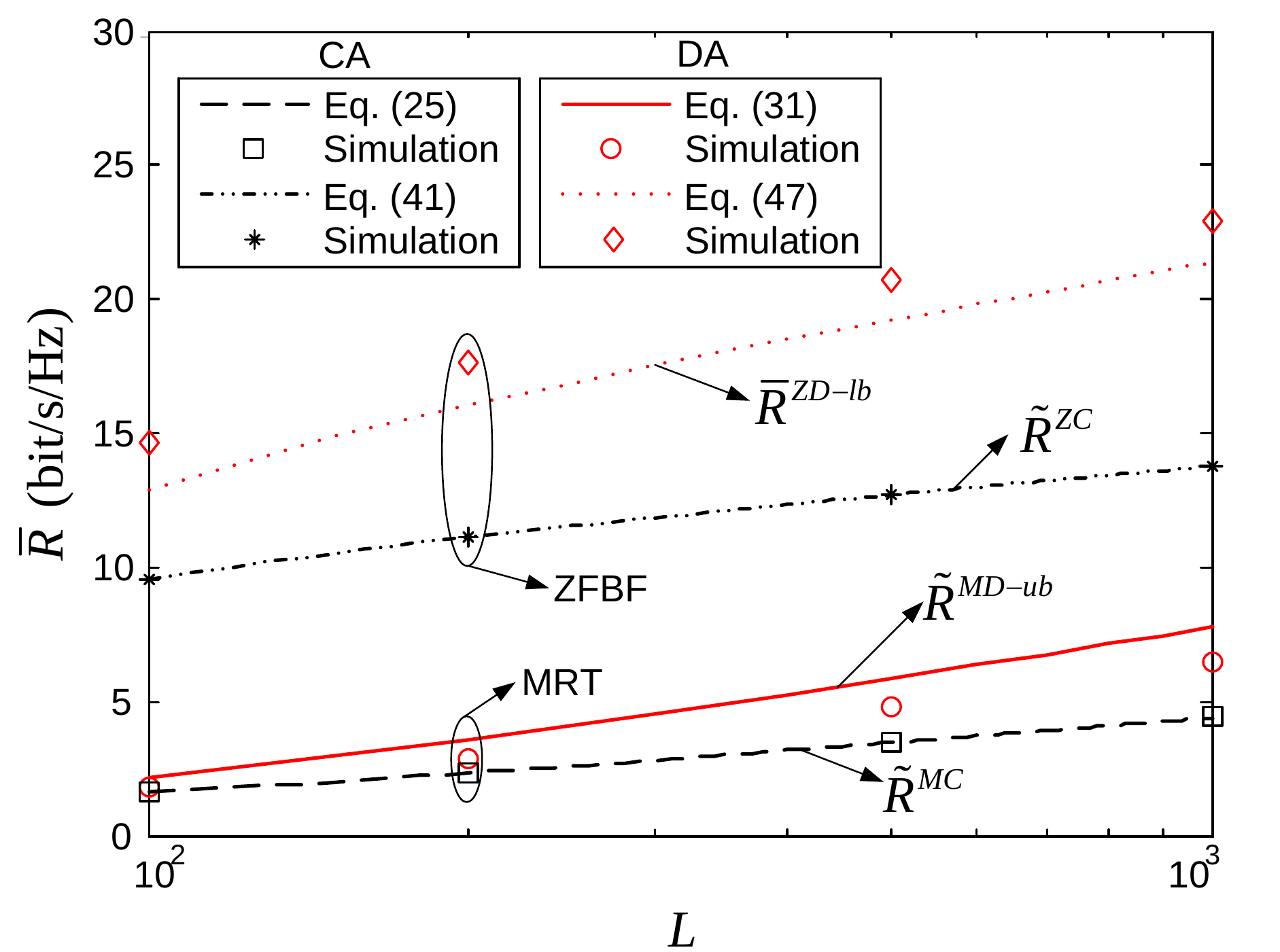}
\caption{Average user rate $\bar{R}$ versus the number of BS antennas $L$ with MRT and ZFBF in the CA layout and the DA layout. $P_{t}/N_{0}=20$dB. $\alpha=4$. $K=50$.}
\label{FIG_Rk_FixedK}
\end{center}
\end{figure}

For a better understanding of the effect of precoding schemes on the comparison results, Fig. \ref{FIG_Rk_FixedK} illustrates the average user rates of both MRT and ZFBF when the number of users $K$ is fixed at 50. Recall that users suffer from significant intra-cell interference if MRT is adopted. By spreading out the BS antennas, both the desired signal power and the intra-cell interference are enhanced, and thus only marginal gains over the CA layout can be observed. With ZFBF, in contrast, the intra-cell interference is eliminated by joint precoding over users. Substantial rate gains can be then achieved from the reduction of the minimum access distance as shown in Fig. \ref{FIG_Rk_ZFBF}. We can then conclude that although a higher average user rate is always achieved by the DA layout in both MRT and ZFBF cases, the gains are much more prominent if an orthogonal precoding scheme is adopted.

\section{Conclusion}

This paper presents a comparative study of the asymptotic rate performance of downlink multi-user systems with two BS antenna layouts, i.e., the CA layout and the DA layout. By assuming that the number of BS antennas $L$ and the number of users $K$ grow infinitely while $L/K{\rightarrow} \upsilon$, simple explicit expressions of the asymptotic average user rate with the CA layout under two representative linear precoding schemes, MRT and ZFBF, are derived and shown to be good approximations for the finite case. For the DA layout, bounds are developed to study the scaling behavior of the rate performance.

The analysis shows that with the CA layout, the asymptotic average user rates for both MRT and ZFBF are logarithmic functions of the ratio $\upsilon$. With the DA layout, in contrast, the scaling behavior of the average user rate closely depends on the precoding schemes. With MRT, an asymptotic upper-bound of the average user rate with the DA layout is obtained, which also logarithmically increases with $\upsilon$, but at a higher growth rate than that in the CA case. With ZFBF, a lower-bound is developed and found to be increasing unboundedly as $L, K{\rightarrow} \infty$ and $L/K{\rightarrow} \upsilon{>} 1$, implying that the average user rate with the DA layout is much higher than that with the CA layout when the number of BS antennas is large. Simulation results corroborate that the bounds well indicate how the average user rate scales with the number of BS antennas $L$. For large $L$, the DA layout has better rate performance in both MRT and ZFBF cases, and more substantial rate gains over the CA layout are achieved when the orthogonal precoding scheme, ZFBF, is adopted. The analysis provides direct guidance to the cellular system design, and serves as a benchmark for the rate analysis with more sophisticated precoding schemes in the future.

Note that in this paper, the total transmission power is assumed to be fixed for the sake of comparison between the CA layout and the DA layout. For the DA layout, however, each distributed BS antenna may have its own power constraint. It is therefore important to further characterize the downlink rate performance with a large number of distributed BS antennas under the per-antenna power constraint. Moreover, the linear precoding schemes considered in this paper, MRT and ZFBF, both require joint transmission among all the BS antennas, which may not be necessary for the DA layout. In this case, each user is close to a very limited number of BS antennas and thus only a small subset of BS antennas may need to be considered for each user's transmission. How to design efficient transmission schemes for the distributed antenna systems is an interesting topic, which deserves much attention in the future study.

\appendices

\section{Derivation of (\ref{mu_MD})}
The average received SINR $\mu_{k}^{MD}$ is determined by $a_{j,l}$ according to (\ref{mu_M}). It can be easily obtained from (\ref{define of al}) that
\begin{align}\label{al_MD_1}
a_{j,l}&{=}\mathbb{E}_{\mathbf{h}_{j}}\left[\frac{|\tilde{g}_{j,l}|^2}{\|\tilde{\mathbf{g}}_{j}\|^2}\right]
=\mathbb{E}_{\mathbf{h}_{j}}\left[\frac{\beta_{j,l}^{2}|h_{j,l}|^2}
{\beta_{j,l}^{2}|h_{j,l}|^2+\sum_{m\in \mathcal{B}, m\neq l}|\tilde{g}_{j,m}|^2}\right] \nonumber \\
&{=}\mathbb{E}_{x,y}\left[\frac{\beta_{j,l}^{2}y}{\beta_{j,l}^{2}y+x}\right],
\end{align}
where $x=\sum_{m\in \mathcal{B}, m\neq l}|\tilde{g}_{j,m}|^2$ is a hypoexponential random variable with the probability density function (pdf) \cite{Ross}
\begin{equation}\label{pdf of x}
f(x){=}\hspace{-0.3cm}\sum_{m\in\mathcal{B}, m\neq l}\hspace{-0.2cm}\beta_{j,m}^{-2}\exp\left\{-\beta_{j,m}^{-2}x\right\}
\hspace{-0.3cm}\prod_{t\in \mathcal{B}, t\neq m, t\neq l}\hspace{-0.1cm}\frac{\beta_{j,t}^{-2}}{\beta_{j,t}^{-2}{-}\beta_{j,m}^{-2}},
\end{equation}
and $y=|h_{j,l}|^2$ is an exponential random variable with the pdf
\begin{equation}\label{pdf of y}
f(y)=\exp\{-y\},
\end{equation}
which is independent of $x$.
By combining (\ref{al_MD_1}-\ref{pdf of y}), we have
\begin{align}\label{al_MD_2}
a_{j,l}&=\int_{0}^{\infty}\int_{0}^{\infty}\frac{\beta_{j,l}^{2}y}{\beta_{j,l}^{2}y+x}\left(\sum_{m\in\mathcal{B}, m\neq l}\beta_{j,m}^{-2}
\exp\left\{-\beta_{j,m}^{-2}x\right\}\prod_{t\in \mathcal{B}, t\neq m, t\neq l}\frac{\beta_{j,t}^{-2}}{\beta_{j,t}^{-2}-\beta_{j,m}^{-2}}\right)dx \cdot \exp\{-y\}dy  \nonumber \\
&\mathop{=}\limits^{u=\beta_{j,l}^{2}y+x}\hspace{-0.4cm}\sum_{m\in\mathcal{B}, m\neq l}\hspace{-0.3cm}\beta_{j,m}^{-2}\hspace{-0.1cm}
\left(\hspace{-0.1cm}\int_{0}^{\infty}\hspace{-0.2cm}\beta_{j,l}^{2}y\exp\left\{\beta_{j,l}^{2}\beta_{j,m}^{-2}y\right\}\hspace{-0.2cm}
\int_{\beta_{j,l}^{2}y}^{\infty}\hspace{-0.2cm}u^{-1}\hspace{-0.1cm}\exp\{-\beta_{j,m}^{-2}u\}du\cdot \exp\{-y\}dy\hspace{-0.1cm}\right)\hspace{-0.2cm}
\prod_{t\in \mathcal{B}, t\neq m, t\neq l}\frac{\beta_{j,t}^{-2}}{\beta_{j,t}^{-2}-\beta_{j,m}^{-2}} \nonumber \\
&\mathop{=}\limits^{z=(\beta_{j,l}^2y)^{-1}u}\hspace{-0.4cm}\sum_{m\in\mathcal{B}, m\neq l}\beta_{j,l}^{2}\beta_{j,m}^{-2}
\left(\int_{0}^{\infty}y\exp\left\{\beta_{j,l}^{2}\beta_{j,m}^{-2}y\right\}
\hspace{-0.1cm}\int_{1}^{\infty}\hspace{-0.2cm}z^{-1}\exp\left\{-\beta_{j,l}^{2}\beta_{j,m}^{-2}yz\right\}dz
\cdot \exp\{-y\}dy\right) \nonumber \\
&\prod_{t\in \mathcal{B}, t\neq m, t\neq l}\frac{\beta_{j,t}^{-2}}{\beta_{j,t}^{-2}-\beta_{j,m}^{-2}} \nonumber \\
&=\sum_{m\in\mathcal{B}, m\neq l}\beta_{j,l}^{2}\beta_{j,m}^{-2}
\left(\int_{1}^{\infty}z^{-1}\int_{0}^{\infty}y\exp\left\{\left[\beta_{j,l}^{2}\beta_{j,m}^{-2}(1-z)-1\right]y\right\} dydz \right)
\prod_{t\in \mathcal{B}, t\neq m, t\neq l}\frac{\beta_{j,t}^{-2}}{\beta_{j,t}^{-2}-\beta_{j,m}^{-2}}\nonumber \\
&=\sum_{m\in\mathcal{B}, m\neq l}\frac{\beta_{j,l}^{-2}\beta_{j,m}^{-2}(\log\beta_{j,l}^{-2}-\log\beta_{j,m}^{-2}-1)+\beta_{j,m}^{-4}}
{(\beta_{j,l}^{-2}-\beta_{j,m}^{-2})^{2}}\prod_{t\in \mathcal{B}, t\neq m, t\neq l}\frac{\beta_{j,t}^{-2}}{\beta_{j,t}^{-2}-\beta_{j,m}^{-2}}.
\end{align}
(\ref{mu_MD}) can be then obtained by substituting (\ref{al_MD_2}) into (\ref{mu_M}).

\section{Derivation of (\ref{AsymR_MD_ub})}
Let us first focus on the asymptotic behavior of the SINR upper-bound $\mu_{k}^{MD-ub}$. According to (\ref{mu_MD_ub}), $\mu_k^{MD-ub}$ is a function of 1) $m_k$, the number of interfering users whose closest antenna is the same as user $k$'s, and 2) $Y=d_{k}^{j_{(1)}}/{d_{k}^{l_{(1)}}}$, where  $d_{k}^{l_{(1)}}$ and $d_{k}^{j_{(1)}}$ denote the minimum access distances from user $k$ to $L$ BS antennas and the other $K-1$ users, respectively. With $L$ BS antennas uniformly distributed in a disk with radius $1$, it has been shown in \cite{Dai_JSAC} that the minimum access distance $d_{k}^{l_{(1)}}$ from user $k$ to $L$ BS antennas given user $k$'s position at $(\rho_k,\theta_k)$ follows the pdf
\setcounter{equation}{52}
\begin{equation}\label{pdf of dmin}
f_{d_{k}^{l_{(1)}}|\rho_{k}}(x|y)=L\left(1-F(x; y)\right)^{L-1}f(x; y),
\end{equation}
where $f(x;y)$ and $F(x; y)$ are given in (\ref{pdf of dl}) and (\ref{cdf of dl}), respectively. Similarly, with $K-1$ users uniformly distributed in a disk with radius 1, the conditional pdf of the minimum access distance $d_{k}^{j_{(1)}}$ from user $k$ to $K-1$ users given user $k$'s position at $(\rho_{k}, \theta_{k})$ can be written as
\begin{equation}\label{pdf of djmin}
f_{d_{k}^{j_{(1)}}|\rho_{k}}(x|y)=\left(K-1\right)\left(1-F(x;y)\right)^{K-2}f(x;y).
\end{equation}
By combining (\ref{pdf of dmin}) and (\ref{pdf of djmin}), the conditional pdf of $Y=d_{k}^{j_{(1)}}/{d_{k}^{l_{(1)}}}$ given user $k$'s position at $(\rho_{k}, \theta_{k})$ can be obtained as
\begin{align}\label{pdf_Y}
f_{Y|\rho_{k}}\left(y|t\right)&=\int_{0}^{1+t}f_{d_{k}^{j_{(1)}}|\rho_{k}}\left(xy|t\right)
\cdot f_{d_{k}^{l_{(1)}}|\rho_{k}}\left(x|t\right)\cdot x dx \nonumber \\
&\mathop{\approx}\limits^{{\rm for\; large\; }L,\;K}  \hspace{-0.1cm} \int_{0}^{1{-}t}\left(K{-}1\right)\exp\left\{{-}\left(K{-}2\right)x^{2}y^{2}\right\} \nonumber \\
&\cdot 2xy\cdot L\exp\left\{-\left(L-1\right)x^{2}\right\}\cdot 2x \cdot x dx \nonumber \\
&=\hspace{-0.1cm}\frac{2yL\left(K{-}1\right)}{\left(L{-}1{+}\left(K{-}2\right)y^{2}\right)^{2}}\hspace{-0.1cm}\left[1{-}\hspace{-0.1cm}\left(1\hspace{-0.1cm}
+\hspace{-0.1cm}\left(L{-}1\hspace{-0.1cm}+\hspace{-0.1cm}\left(K{-}2\right)y^{2}\right)\right. \right.\nonumber \\
&\left.\left.\hspace{-0.1cm}(1{-}t)^{2}\right)\hspace{-0.1cm}
\exp\hspace{-0.1cm}\left\{\hspace{-0.1cm}-\hspace{-0.1cm}\left(L{-}1{+}\left(K{-}2\right)y^{2}\right)\hspace{-0.1cm}(1{-}t)^{2}\right\}\right].
\end{align}
As $L, K{\rightarrow} \infty$ and $L/K{\rightarrow} \upsilon$, $\left(1{+}\left(L{-}1{+}\left(K{-}2\right)y^{2}\right)(1{-}t)^{2}\right)$ $\exp\left\{{-}\left(L{-}1{+}\left(K{-}2\right)y^{2}\right)(1{-}t)^{2}\right\}{\rightarrow} 0$. Then we have
\begin{equation}\label{pdf_Y_Asym}
f_{Y}\left(y\right)\rightarrow\frac{2\upsilon y}{\left(\upsilon+y^2\right)^{2}}.
\end{equation}
On the other hand,
with $L$ uniformly distributed BS antennas, $m_{k}$ follows the binomial distribution with parameters $K-1$ and $1/L$. As $L, K\rightarrow \infty$ and $L/K \rightarrow \upsilon$, it converges to the Poisson distribution with parameter $1/\upsilon$:
\begin{equation}\label{pdf of m_k}
\text{Pr}\left\{m_{k}=n\right\}=\frac{\upsilon^{-n}e^{-\frac{1}{\upsilon}}}{n!},
\end{equation}
$n=1, 2, ...$. By combining (\ref{Rk_MD_approx}) and (\ref{mu_MD_ub}), we have
\begin{equation}\label{Rk_MD_ub}
R_{k}^{MD}{\leq} R_{k}^{MD-ub}{=}\hspace{-0.1cm}\left\{ {\begin{array}{*{20}{c}}
{\hspace{-0.2cm}\exp\{m_{k}\}E_{1}\{m_{k}\}\log_{2}e}&{\hspace{-0.2cm}\text{if}\quad{m_k} \ne 0,}\\
{\hspace{-0.2cm}\exp\{Y^{-\alpha}\}E_{1}\{Y^{-\alpha}\}\log_{2}e}&{\hspace{-0.2cm}\text{otherwise}.}
\end{array}} \right.
\end{equation}
As $L, K\rightarrow \infty$ and $L/K\rightarrow \upsilon$, $\bar{R}^{MD-ub}$ converges to
\begin{align}\label{AverageR_MD_ub_ap}
\bar{R}^{MD-ub}&{\rightarrow}\left(\sum_{n=1}^{\infty}\exp\left\{n\right\}\hspace{-0.1cm}E_{1}\left\{n\right\} \frac{\upsilon^{-n}e^{-\frac{1}{\upsilon}}}{n!} {+}\hspace{-0.2cm}\int_{0}^{\infty}\hspace{-0.2cm}\exp\left\{y^{-\alpha}\hspace{-0.1cm}{-}\frac{1}{\upsilon}\right\} \right. \nonumber \\
&\left.\hspace{-0.1cm}E_{1}\left\{y^{-\alpha}\right\} \frac{2\upsilon y}{(\upsilon+y^2)^2}dy\right)\hspace{-0.1cm}\log_{2}e,
\end{align}
according to (\ref{AverageR_MD_ub}) and (\ref{pdf_Y_Asym}-\ref{Rk_MD_ub}). (\ref{AsymR_MD_ub}) can be then obtained from (\ref{AverageR_MD_ub_ap}).

\section{Derivation of (\ref{Rk_ZD_lb})}
It is shown in (\ref{Rk_mu_Z}) that the maximum achievable ergodic rate of user $k$ with ZFBF is determined by the the effective channel gain $1/{\|\mathbf{f}_{k}\|^{2}}$. With a large number of BS antennas $L$, each user $j$ is close to some BS antenna $l_{j}^{*}$. With $L\gg K$, the probability that two interfering users $j_{1}$ and  $j_{2}$ are close to the same BS antenna is very low, i.e., $\text{Pr}\left\{l_{j_{1}}^{*}=l_{j_{2}}^{*}|_{j_{1}\neq j_{2}}\right\}\approx0$. As a result, we have $\tilde{\mathbf{g}}_{j_{1}}\tilde{\mathbf{g}}_{j_{2}}^{\dag}=0$ for $j_{1}\neq j_{2}$.
The effective channel gain can be then obtained from (\ref{EffectiveChannelGain}) as
\begin{align}\label{EffectiveChannelGain_ap}
\frac{1}{\|\mathbf{f}_{k}\|^{2}}
&\mathop{\approx}\limits^{L\gg K}\|\tilde{\mathbf{g}}_{k}\|^2-\sum_{j\neq k, j\in\mathcal{K}}\tilde{\mathbf{g}}_{k}\tilde{\mathbf{g}}_{j}^{\dag}\cdot \frac{1}{\|\tilde{\mathbf{g}}_{j}\|^{2}}\cdot \tilde{\mathbf{g}}_{j}\tilde{\mathbf{g}}_{k}^{\dag} \nonumber \\
&=\|\tilde{\mathbf{g}}_{k}\|^2-\sum_{j\neq k, j\in\mathcal{K}}|\tilde{g}_{k,l_{j}^{*}}|^{2}
=\sum_{l\in {\tilde{\mathcal{B}}_{k}}}|\tilde{g}_{k,l}|^{2},
\end{align}
where $\tilde{\mathcal{B}}_{k}=\mathcal{B}-\left\{l_{j}^{*}\right\}_{j\in\mathcal{K}, j\neq k}$. By combining (\ref{Rk_mu_Z}-\ref{mu_Z}) and (\ref{EffectiveChannelGain_ap}), the maximum achievable ergodic rate  of user $k$ with ZFBF in the DA layout $R_{k}^{ZD}$ can be obtained as
\begin{align}\label{Rk_ZD}
R_{k}^{ZD}&{=}\mathbb{E}_{\mathbf{h}_{k}}\hspace{-0.2cm}\left[\log_{2}\left(1+\frac{\bar{P}_{k}}{N_{0}} \sum_{l\in \tilde{\mathcal{B}}_{k}}|g_{k,l}|^{2}\right)\right] \nonumber \\
&{=}\mathbb{E}_{\mathbf{h}_{k}}\hspace{-0.2cm}\left[\log_{2}\hspace{-0.1cm}\left(\hspace{-0.1cm}1{+}
\frac{P_{t}}{KN_{0}}\hspace{-0.1cm}\sum_{l\in \tilde{\mathcal{B}}_{k}}\hspace{-0.1cm}\left(d_{k,l}\right)^{-\alpha}|h_{k,l}|^{2}\right)\right],
\end{align}
where $d_{k, l}$ denotes the access distance from user $k$ to BS antenna $l\in \tilde{\mathcal{B}}_{k}$. Since both $L$ BS antennas and $K$ users are uniformly distributed within a circular cell, $\tilde{\mathcal{B}}_{k}$ is composed of $L-K+1$ uniformly distributed BS antennas. Let $\tilde{d}_{k}^{l_{(1)}}=\min_{l\in \tilde{\mathcal{B}}_{k}}d_{k,l}$ and $\tilde{l}_{k}^{*}={\arg\min}_{l\in\tilde{\mathcal{B}}_{k}}d_{k, l}$, we have
\begin{align}\label{Ap}
\sum_{l\in \tilde{\mathcal{B}}_{k}}\left(d_{k,l}\right)^{-\alpha}|h_{k,l}|^{2}
\geq \left(\tilde{d}_{k}^{l_{(1)}}\right)^{-\alpha}|h_{k,\tilde{l}_{k}^{*}}|^{2}.
\end{align}
(\ref{Rk_ZD_lb}) can be then obtained by combining (\ref{Rk_ZD}-\ref{Ap}).

\section{derivation of (\ref{AsymR_ZD})}

According to (\ref{AverageR_ZD_lb_1}), we have
\begin{align}\label{AverageR_ZD_lb_ap}
\bar{R}^{ZD{-}lb}
&\hspace{-0.15cm}\geq \hspace{-0.1cm} \exp\hspace{-0.1cm}\left\{\hspace{-0.1cm}\tfrac{KN_{0}}{P_{t}}\mathbb{E}_{\tilde{d}_{k}^{l_{(1)}}}
\hspace{-0.2cm}\left[\hspace{-0.1cm}\left(\hspace{-0.1cm}\tilde{d}_{k}^{l_{(1)}}\hspace{-0.1cm}\right)^{\alpha}\right]\hspace{-0.1cm}\right\}
\hspace{-0.1cm}E_{1}\hspace{-0.1cm}\left\{\hspace{-0.1cm}\tfrac{KN_{0}}{P_{t}}\mathbb{E}_{\tilde{d}_{k}^{l_{(1)}}}\hspace{-0.2cm}
\left[\hspace{-0.1cm}\left(\hspace{-0.1cm}\tilde{d}_{k}^{l_{(1)}}\hspace{-0.1cm}\right)^{\alpha}\right]\hspace{-0.1cm}\right\}\hspace{-0.1cm}\log_{2}e,
\end{align}
as $f(x)=\exp\{x\}E_{1}\{x\}$ is a convex function. The expectation of $\left(\tilde{d}_{k}^{l_{(1)}}\right)^{\alpha}$ can be written as 
\begin{align}\label{E_dmin'}
\mathbb{E}_{\tilde{d}_{k}^{l_{(1)}}}\left[\left(\tilde{d}_{k}^{l_{(1)}}\right)^{\alpha}\right]
&{=}\mathbb{E}_{\rho_{k}}\left[\mathbb{E}_{\tilde{d}_{k}^{l_{(1)}}|\rho_{k}}
\left[\left(\tilde{d}_{k}^{l_{(1)}}\right)^{\alpha}|\rho_{k}\right]\right],
\end{align}
where the conditional expectation $\mathbb{E}_{\tilde{d}_{k}^{l_{(1)}}|\rho_{k}}\left[\left(\tilde{d}_{k}^{l_{(1)}}\right)^{\alpha}|\rho_{k}\right]$, according to (\ref{pdf of dmin'}), can be obtained as
\begin{align}\label{AsymEdmin}
&\mathbb{E}_{\tilde{d}_{k}^{l_{(1)}}\hspace{-0.1cm}|\rho_{k}}\hspace{-0.1cm}
\left[\left(\tilde{d}_{k}^{l_{(1)}}\right)^{\alpha}|\rho_{k}\right]
{=}\int_{0}^{1+t}x^{\alpha}\cdot f_{\tilde{d}_{k}^{l_{(1)}}|\rho_{k}}(x|t)dx \nonumber \\
&\mathop{{\approx}} \limits^{L\gg K}\hspace{-0.2cm} \int_{0}^{1-t}\hspace{-0.2cm}x^{\alpha}\hspace{-0.1cm}\cdot \hspace{-0.1cm}(L{-}K{+}1)
\exp\left\{{-}(L{-}K)x^{2}\right\}\hspace{-0.1cm}\cdot \hspace{-0.1cm}2x dx \nonumber \\
&{=}\frac{L{-}K{+}1}{(L{-}K)^{1{+}\frac{\alpha}{2}}}\hspace{-0.1cm}\left(\Gamma\hspace{-0.1cm}\left(1{+}\tfrac{\alpha}{2}, 0\right)\hspace{-0.1cm}-\hspace{-0.1cm}\Gamma\hspace{-0.1cm}\left(1{+}\tfrac{\alpha}{2}, (L{-}K)(1{-}t)^{2}\right)\right),
\end{align}
where $\Gamma(s,x)\hspace{-0.1cm}=\hspace{-0.1cm}\int_{x}^{\infty}\hspace{-0.1cm}t^{s{-}1}e^{{-}t}dt$. As $L, K{\rightarrow} \infty$ and $L/K {\rightarrow }\upsilon {>}1$, 
$\Gamma(1+\frac{\alpha}{2}, (L-K)(1-t)^{2}){\rightarrow} 0$ and $(L-K+1)/(L-K)^{1+\frac{\alpha}{2}}{\rightarrow} 0$ with the path-loss factor $\alpha>2$. Then we have
\begin{equation}\label{AsyEdminSquare}
\mathbb{E}_{\tilde{d}_{k}^{l_{(1)}}|\rho_{k}}\left[\left(\tilde{d}_{k}^{l_{(1)}}\right)^{\alpha}|\rho_{k}\right]\rightarrow 0.
\end{equation}
Finally, (\ref{AsymR_ZD}) can be obtained by combining (\ref{AverageR_ZD_lb_ap}-\ref{E_dmin'}) and (\ref{AsyEdminSquare}).

\end{document}